\documentclass[10pt]{article}
\usepackage{amssymb}
\usepackage{amsmath}
\usepackage{amsthm}
\usepackage{latexsym}
\usepackage[dvips]{epsfig}
\usepackage{mathrsfs}
\usepackage{eufrak}

\theoremstyle{plain}

\newtheorem{lemma}{Lemma}
\newtheorem{theorem}{Theorem}

\font\SYM=msbm10
\newcommand{\Real}{\mbox{\SYM R}}
\newcommand{\Complex}{\mbox{\SYM C}}
\newcommand{\Natural}{\mbox{\SYM N}}

\font\tenscr=rsfs10 scaled1100
\font\sevenscr=rsfs7 
\font\fivescr=rsfs5 
\skewchar\tenscr='177
\skewchar\sevenscr='177
\skewchar\fivescr='177
\newfam\scrfam
\textfont\scrfam=\tenscr
\scriptfont\scrfam=\sevenscr
\scriptscriptfont\scrfam=\fivescr

\def\scri{{\fam\scrfam I}}

\newcommand{\TT}[3]{T_{#1 \phantom{#2} #3}^{\phantom{#1} #2}}
\newcommand{\updn}[3]{#1^{#2}_{\phantom{#2}#3}}
\newcommand{\dnup}[3]{#1_{#2}^{\phantom{#2}#3}}

\begin{document}


\title{\textbf{Estimates for the Maxwell field near the spatial and null infinity of the Schwarzschild spacetime}}

\author{{\Large Juan Antonio Valiente Kroon} \thanks{E-mail address:
 {\tt j.a.valiente-kroon@qmul.ac.uk}} \\
School of Mathematical Sciences,\\ Queen Mary, University of London,\\
Mile End Road, London E1 4NS,
\\United Kingdom.}

\maketitle

\begin{abstract}
  It is shown how the gauge of the ``regular finite initial
  value problem at spacelike infinity'' can be used to construct a
  certain type of estimates for the Maxwell field propagating on a
  Schwarzschild background. These estimates are constructed with the
  objective of obtaining information about the smoothness near
  spacelike and null infinity of a wide class of solutions to the
  Maxwell equations.
\end{abstract}



\section{Introduction}
In reference \cite{Val07b}, an analysis of the behaviour of the
Maxwell field (spin-1 zero-rest-mass field) propagating near the
spatial infinity of a Schwarzschild background was suggested as a way
of gaining insight into certain aspects of the asymptotics of the
gravitational field. Using Friedrich's \emph{cylinder at spatial
infinity} representation of the region of the Schwarzschild spacetime
which is ``near'' spatial infinity \cite{Fri98a,Fri04} it was possible
to discuss the occurrence of obstructions to the smoothness of the
Maxwell field at null infinity. The analysis of these obstructions is
done by means of a certain type of asymptotic expansions which can be
calculated in the aforementioned formalism. The test Maxwell field
propagating on the is obtained as the solution to an initial value
problem with initial data prescribed on a $t=\mbox{\emph{constant}}$
slice of the conformally rescaled Schwarzschild spacetime. The most
important aspect of the expansions is that they allow to relate in an
explicit manner properties of the initial data with the behaviour of
the field at null infinity. In particular, it was shown in
\cite{Val07b} that these asymptotic expansions contain logarithmic
divergences at the sets where spatial infinity ``touches'' null
infinity ---the so-called \emph{critical sets}. A certain subset of
the logarithmic divergences is still present if instead of
propagation on a Schwarzschild background, one considers propagation
on a flat background ---hence, one can regard these logarithmic
divergences as a structural property of the class of
hyperbolic equations under consideration. This type of logarithmic
divergences was first observed in the analysis of the conformal field
equations carried out in \cite{Fri98a}. As in the case of the conformal
field equations, the analogous logarithmic singularities in the
Maxwell field can be precluded by imposing certain regularity
conditions on the initial data. The analysis in \cite{Val07b} shows
that even if these regularity conditions are satisfied, there are some
further logarithmic divergences which could be interpreted as arising
from the interaction of the Maxwell field with the curved
background. These logarithmic divergences are similar in structure to
the ones observed in \cite{Val04a,Val04d,Val04e,Val05a}.

Due to the hyperbolic nature of the Maxwell equations it is to be
expected that the logarithmic divergences in the asymptotic expansions
will propagate into null infinity, and hence will have an effect on
the smoothness of the test field at the conformal boundary. In view of
the results of \cite{Val07b}, the challenge is to determine in a
precise and rigorous manner how these properties of the asymptotic
expansions translate into properties of the actual solutions to the
Maxwell equations.

The fundamental structural properties of the type of evolution equations
under consideration have been analysed at length elsewhere ---see
\cite{Fri98a,Fri03a,Fri03b,Fri04,Val03a} in the case of the conformal field
equations and \cite{Val07b} in the case of the Maxwell equations. The crucial
observation message in these analyses is that although the relevant
propagation equations happen to be symmetric-hyperbolic in the
interior of the conformally rescaled spacetime, they degenerate at the
critical sets in the sense that the matrix associated with the time
derivative looses rank ---this degeneracy is responsible, in
particular, of the first class of logarithmic singularities discussed
in previous paragraphs. As a consequence, the standard existence
arguments for symmetric hyperbolic systems break down at the critical
sets.

Gaining control on solutions of degenerate propagation equations
requires an understanding of their \emph{algebraic} properties ---this
is the rationale of the analysis in \cite{Val07b}. In a second stage,
ideally, one has to learn how to relate these algebraic properties
with estimates of the solutions to the equations. A first step in this
direction has been given by Friedrich in \cite{Fri03b} with the
construction of a certain type of estimates for the spin-2 field on
the Minkowski spacetime. This construction bypasses the technical
problems mentioned in the previous paragraphs, and permits the use of
information coming from asymptotic expansions.

The objective of the present work is to adapt the ideas of
\cite{Fri03b} to the case of the Maxwell field on a curved background
---the Schwarzschild spacetime. The implementation these ideas is far from straightforward. Some of
the problems one faces in this implementation were already foreseen in
\cite{Fri03b}: the construction of estimates was performed in a very
particular conformal gauge, which in a sense, exploits to the maximum
the simplicity of the Minkowski spacetime. The use of other gauges
would include a string of extra terms which one has to learn how to
control. As we shall see, the situation in a curved background is
analogous: the curved background also introduces a string of lower order terms. The crucial idea in the
construction of estimates for the spin-2 massless field on a flat
background is that although at first
sight it seems not possible to construct $L^2$-type estimates for the
components of the field, it is nevertheless possible by means of an
alternative argument to construct estimates for sufficiently high
``radial'' derivatives. These estimates, in turn, can be used to
control the remainder of Taylor-like expansions of the Maxwell field.
In order to extend these ideas to the case of a curved background, it
will be necessary to take this insight a step further and assume that
one has solutions to the Maxwell equations have a Taylor-like expansion
---this last point was not assumed in \cite{Fri03b}.

\bigskip This article is organised as follows: section \ref{F_gauge}
gives a brief review of Friedrich's formalism of the cylinder at
spatial infinity applied to the case of the Schwarzschild spacetime.
This will be the setting of our discussion. Section
\ref{section:Maxwell} gives a discussion of relevant aspects of the
Maxwell equations within the framework of the cylinder at spatial
infinity. Section \ref{section:Minkowski} contains a discussion of the
construction of estimates for the Maxwell field on a Minkowski
background. This section follows very closely a similar discussion for
the spin-2 massless field given in \cite{Fri03b}, and is given here
for completeness, reference and comparison with the discussion in
subsequent sections. Section \ref{section:Schwarzschild} gives the
construction of estimates on a Schwarzschild background, and contains
the main results of the article. Finally, there is a concluding
section ---section \ref{section:conclusions}--- containing a summary
of the results obtained and listing all the assumptions being made. In
addition there are two appendices: the first one, appendix
\ref{appendix:spinors}, listing the definitions of some spinorial
objects used throughout the article, and a second one, appendix
\ref{appendix:SU(2)}, describing some crucial results concerning vector
fields on the Lie algebra of $SU(2)$.

\section{The Schwarzschild spacetime in the F-gauge} \label{F_gauge}
Some relevant aspects of the framework of the \emph{cylinder at
spatial infinity} or \emph{F-gauge} for the case of the Schwarzschild
spacetime are first discussed. This gauge allows to formulate, for the
Einstein equations, a \emph{regular finite initial value problem near
spatial infinity}. The original construction has been given in
\cite{Fri98a}. The Schwarzschild metric in
isotropic coordinates is given by 
\[
\tilde{g}=\left(\frac{1-m/2r}{1+m/2r}\right)^2\mbox{d}t^2
+\left(1+\frac{m}{2r} \right)^4 \big( \mbox{d}r^2 + r^2
\mbox{d}\sigma^2\big).
\]
Consider the time-symmetric hypersurface $\tilde{\mathcal{S}}$,
$t=\mbox{\emph{constant}}$ and let $\tilde{h}$ denote its (negative
definite) intrinsic metric.  Writing $\tilde{h}=\Omega^{-2}h$, one
finds the following conformal intrinsic metric and conformal factor:
\begin{equation}
h=-\big( \mbox{d}\rho^2 + \rho^2 \mbox{d}\sigma^2 \big), \quad \Omega=\frac{\rho^2}{(1+\rho m/2)^2}, \label{Omega}
\end{equation}
where the radial coordinate $\rho=1/r$ has been introduced. Let $i_1$
and $i_2$ denote the infinities corresponding to the two asymptotic
ends of the hypersurface $\tilde{\mathcal{S}}$. Further, let
$\mathcal{S}=\tilde{\mathcal{S}}\cup\{i_1,i_2 \}$. Let $i=i_1$ denote
the infinity corresponding to the locus $\rho=0$. The discussion in
this article will be concerned with the domain of influence,
$J^+(\mathcal{B}_{a_*}(i))$ of a sufficiently small ball,
$\mathcal{B}_a(i)$, of radius $a_*$ based on $i$. The point $i$ can be
blown up to a 2-sphere, $\mathcal{S}^2$. Accordingly, introduce the
set $\mathcal{C}_{a_*}=(\mathcal{B}_{a_*}(i)\setminus {i})\cup \mathcal{S}^2$.

The use of a gauge based on \emph{conformal Gaussian coordinates}
leads to a spacetime conformal factor given by
\begin{equation}
\Theta = \frac{\Omega}{\kappa}(1-\tau^2\frac{\kappa^2}{\omega^2}), \label{cf:Theta}
\end{equation}
where
\begin{equation}
\omega=\frac{2\Omega}{\sqrt{|D^i\Omega D_i\Omega|}}=\rho(1+\rho m/2), \label{omega}
\end{equation}
and $\kappa>0$ is a smooth function such that
$\kappa=\mu\rho$ with $\mu(i)=1$; $D_i$ denotes
the Levi-Civita connection of the (flat) metric $h$. The function $\kappa$
encodes the remaining conformal freedom in the setting. In order to
avoid disrupting the spherical symmetry of the representation, the
analysis will be restricted to spherically symmetric choices of
$\kappa$. Usual choices for $\kappa$ are
\begin{subequations}
\begin{eqnarray}
&& \kappa=\rho \label{kappa_rho}, \\
&& \kappa=\omega \label{kappa_omega}.
\end{eqnarray}
\end{subequations}
\emph{For the purposes of the current article, it turns out that the choice
of $\kappa$ given by (\ref{kappa_omega}) is more convenient.}

The coordinate $\tau$ in (\ref{cf:Theta}) is an affine parameter of
conformal geodesics whose tangent at $\tau=0$ is parallel to the
normal of $\mathcal{S}$. Using these conformal geodesics one
constructs \emph{conformal Gaussian coordinates}: the coordinate
$\rho$ can be extended off $\mathcal{S}$ by requiring it to be
constant along the aforementioned conformal geodesics. ``Angular
coordinates'' can be extended in a similar fashion.

In view of the conformal factor (\ref{cf:Theta}) define the manifold
\[
\mathcal{M}_{{a_*},\kappa} =\left\{ (\tau,q)\;|\; q\in \mathcal{C}_{a_*}, \; -\frac{\omega}{\kappa}\leq \tau \leq \frac{\omega}{\kappa} \right\},
\]
and the following relevant subsets thereof:
\begin{subequations}
\begin{eqnarray*}
&& \mathcal{I}=\left\{(\tau,q)\in \mathcal{M}_{{a_*},\kappa} \;|\; \rho(q)=0, |\tau|<\frac{\omega}{\kappa} \right\}, \\
&& \mathcal{I}^\pm = \left\{(\tau,q)\in \mathcal{M}_{{a_*},\kappa} \;|\; \rho(q)=0, \tau=\pm 1\right\}, \\
&& \mathcal{I}^0 =\left\{(\tau,q)\in \mathcal{M}_{{a_*},\kappa} \;|\; \rho=0, \;\tau=0  \right\}, \\
&& \mathscr{I}^\pm =\left\{ (\tau,q)\in \mathcal{M}_{{a_*},\kappa} \;|\; q\in \mathcal{B}_{a_*}(i)\setminus \mathcal{I}^0, \;\tau=\pm\frac{\omega}{\kappa} \right\}, 
\end{eqnarray*}
\end{subequations}
denoting, respectively, the \emph{cylinder at spatial infinity}, the
\emph{critical sets} where spatial infinity touches null infinity, the
intersection of the cylinder at spatial infinity with the initial
hypersurface $\mathcal{S}$, and the two components of \emph{null
  infinity}. In particular, with the choice (\ref{kappa_omega}) of
$\kappa$, the locus of null infinity is given by $\tau=\pm 1$. The manifold
$\mathcal{M}_{{a_*},\kappa}$ with the gauge choice $\kappa=\omega$, will be
denoted by $\mathcal{M}_{{a_*},\omega}$. 

\bigskip
It will be convenient to work with a space-spinor formalism ---see
e.g. \cite{Som80}. In order to write down the field equations, introduce a
null frame $c_{AA'}$ satisfying
$g(c_{AA'},c_{BB'})=\epsilon_{AB}\epsilon_{A'B'}$. Let $\tau_{AA'}$
---with normalisation $\tau^{AA'}\tau_{AA'}=2$--- be tangent to the
conformal geodesics of which $\tau$ is a parameter. The frame can be
split into 
\[
c_{AA'}=\frac{1}{2}\tau_{AA'} \tau^{CC'}c_{CC'}-\updn{\tau}{B}{A'}c_{AB},
\]
with
\[
\tau^{AA'}c_{AA'}=\sqrt{2}\partial_{\tau}, \quad c_{AB}=\dnup{\tau}{(A}{B'}c_{B)B'}.
\]
In particular, the following choice will be made:
\[
c_{00'}=\frac{1}{\sqrt{2}}\left( (1+c^0)\partial_\tau+c^1\partial_\rho\right), \quad c_{11'}=\frac{1}{\sqrt{2}}\left( (1-c^0)\partial_\tau -c^1\partial_\rho \right),
\]
with $c^0$ and $c^1$ functions of $(\tau,\rho)$. The remaining vectors
of the frame, $c_{01'}$ and $c_{10'}$ must then be tangent to the
spheres $\{\tau=\mbox{constant}, \; \rho=\mbox{constant}\}$, and thus
cannot define smooth vector fields everywhere. To avoid this
difficulty all possible tangent vectors $c_{01'}$ and $c_{10'}$ will
be considered. This results in a 5-dimensional submanifold of
the bundle of normalised spin frames. Rotations $c_{01'}\rightarrow
\mbox{e}^{\mbox{i}\vartheta}c_{01'}$, $\vartheta\in\Real$ leave this
submanifold invariant. Hence it defines a subbundle with structure
group $U(1)$ which projects into $\mathcal{M}_{{a_*},\omega}$. All the
relevant structures will be lifted to the subbundle, which in an abuse
of notation will be again denoted by $\mathcal{M}_{{a_*},\omega}$ and which is
diffeomorphic to $[-1,1]\times [0,\infty) \times SU(2)$. 

The introduction of the bundle space $\mathcal{M}_{{a_*},\omega}$ and
of the frame $c_{AA'}$ in our formalism implies that all relevant
quantities have a definite spin weight and hence admit an expansion in
terms of some functions $\TT{j}{k}{l}$ associated with unitary
representations of $SU(2)$ ---see e.g. \cite{Fri98a,Fri04} for a more
detailed discussion in this respect. One can introduce differential
operators $X$, $X_+$ and $X_-$ defined by their action on the
functions $\TT{j}{k}{l}$. With the help of these operators one can
write
\[
c_{AA'}=c^\mu_{AA'}\partial_\mu=c^0_{AA'}\partial_\tau + c^1_{AA'}\partial_\rho+ c^+_{AA'}X_+ +c^-_{AA'}X_-.
\]

In addition to the frame $c_{AA'}$, in the F-gauge the geometry of
$\mathcal{M}_{a,\kappa}$ is described by means of the associated
connection $\Gamma_{AA'BC}$, the spinorial counterpart of the Ricci
tensor of a Weyl connection, $\Theta_{AA'BB'}$, and the rescaled Weyl
spinor, $\phi_{ABCD}$. Its unprimed (i.e. space-spinor) version of the
connection spinor is given by
$\Gamma_{ABCD}=\dnup{\tau}{B}{B'}\Gamma_{AB'CD}$,
$\Gamma_{AA'CD}=\Gamma_{ABCD}\updn{\tau}{B}{A'}$, which is decomposed
as
\[
\Gamma_{ABCD}=\frac{1}{\sqrt{2}}\left(\xi_{ABCD}-\chi_{(AB)CD} \right)-\frac{1}{2}\epsilon_{AB}f_{CD}.
\]
Similarly, one considers
$\Theta_{ABCD}=\dnup{\tau}{C}{A'}\dnup{\tau}{D}{B'}\Theta_{AA'BB'}$.
The explicit spherical symmetry of the spacetime justifies the
following Ansatz in terms of irreducible spinors:
\begin{subequations}
\begin{eqnarray*}
&& c^0_{AB}=c^0 x_{AB}, \quad c^1_{AB}=c^1x_{AB}, \quad c^-_{AB}=c^-y_{AB}, \quad c^+_{AB}=c^+z_{AB}, \label{unknown_1}\\
&& f_{AB}=f x_{AB}, \quad \xi_{ABCD}=\xi (\epsilon_{AC}x_{BD} +\epsilon_{BD}x_{AC}),  \label{unknown_2}\\
&& \chi_{(AB)CD}=\chi_2 \epsilon^2_{ABCD} + \chi_h h_{ABCD}, \label{unknown_3}\\
&& \Theta_{ABCD}=\Theta_2 \epsilon^2_{ABCD}+\Theta_h h_{ABCD}+\Theta_x \epsilon_{AB}x_{CD}, \quad \phi_{ABCD}=\phi\epsilon^2_{ABCD}. \label{unknown_4}
\end{eqnarray*}
\end{subequations}
The definitions of the irreducible spinors introduced above is given
in the appendix. The manifest spherical symmetry of this
representation implies that the functions $c^0$, $c^1$, $c^\pm$, $f$,
$\xi$, $\chi_2$, $\chi_h$, $\phi$ have spin-weight 0. Furthermore they
only contain the function $\TT{0}{0}{0}=1$.

The functions $c^0$, $c^1$, $c^\pm$, $f$, $\xi$, $\chi_2$, $\chi_h$,
$\phi$ are determined by solving the conformal propagation equations
discussed in \cite{Fri98a} with the appropriate initial data. The
problem of reconstructing the conformal Schwarzschild solution from
the given data amounts to finding a solution $u=u(\tau,\rho;m)$ of an
initial value problem of the type
\begin{equation}
\partial_\tau u =F(u,\tau,\rho;m), \quad u(0,\rho;m)=u_0(\rho;m), \label{system:Schwarzschild}
\end{equation}
with analytic functions $F$ and $u_0$. The solution with $m=0$
corresponds to a portion of the conformal Minkowski spacetime, in
which the only non-vanishing components of the solution are given by
\[
c^0=-\mu\tau, \quad c^1=\kappa, \quad c^\pm=\mu, \quad f=\kappa^\prime,
\]
where $'$ denotes differentiation with respect to $\rho$ and $\kappa=\mu\rho$.  Since in this case the solution exists for all $\tau$, $\rho\in
\Real$, it can be shown that for a given $m$ there is sufficiently
small $\rho_0$ such that there is an analytic solution to the system
(\ref{system:Schwarzschild}) which extends beyond $\scri$ for $\rho
< \rho_0<a_*$. Hence, if $a_*$ is taken to be small enough, one can
recover the portion of the Schwarzschild spacetime near null and
spatial infinity.

It follows from the above discussion that the coefficients that are
obtained from solving the transport propagation equations on the
cylinder at spatial infinity ---as discussed in
\cite{Fri98a,Val04a}--- correspond to the first terms in the
expansions of the solutions of (\ref{system:Schwarzschild}).

\subsubsection*{Information about null infinity}
In the sequel, it will be necessary to have some more precise
information about the behaviour of the frame spinors $c_{AB}$. One of
the remarkable features of the present conformal setting is that it
allows to obtain information about some field quantities at null
infinity without the need of explicitly solving the field equations. In
\cite{Fri95} it has been shown that for the conformal Gaussian
coordinates the following relation holds:
\begin{equation}
\Theta f_{AB} =d_{AB} -c_{AB}(\Theta),  \label{frame:info}
\end{equation}
where $d_{AB}$ is a space spinor associated to the conformal
factor $\Theta$, but independent of the choice of $\kappa$ ---see
\cite{Fri98a}. It is determined entirely by the initial data on
$\mathcal{S}$. In the case of the Schwarzschild spacetime one has
\begin{equation}
d_{AB} = \frac{2\rho x_{AB}}{(1+\rho m/2)}. \label{d_AB}
\end{equation}

Thus, if $f_{AB}$, $c_{AB}$ and $d_{AB}$ are smooth at points where
$\Theta=0$ ---as it is the case on null infinity--- then from
(\ref{frame:info}) it follows that
\[
d_{AB}=c_{AB}(\Theta).
\]
Now, in particular 
\[
c_{01}(\Theta)= \frac{1}{\sqrt{2}}\left(c^0 \partial_\tau \Theta + c^1 \partial_\rho \Theta  \right).
\]
For the choice $\kappa=\omega$ the conformal factor (\ref{cf:Theta})
takes the form $\Theta =\Omega\omega^{-1}(1-\tau^2)$, so that one computes
\[
c_{01}(\Theta) \big|_{\scri^+} = -\sqrt{2} \frac{\Omega}{\omega} c^0\big|_{\scri^+}.
\]
Hence, using (\ref{Omega}), (\ref{omega}) and (\ref{d_AB}) one concludes that 
\begin{equation}
c^0\big|_{\scri^+}=-1, \label{c0_scri_1}
\end{equation}
if $\kappa=\omega$. An analogous calculation for $\scri^-$ renders
\begin{equation}
c^0\big|_{\scri-}=1. \label{c0_scri_2}
\end{equation}
Consequences of these results with regards to the
Maxwell equations will be discussed in the following section.

\section{The Maxwell field} \label{section:Maxwell} 
The Maxwell field
will be described by means of totally symmetric valence 2 spinor
$\phi_{AB}$, \emph{the Maxwell spinor}, related to the spinorial
counterpart of the Maxwell tensor via
\[
F_{AA'BB'}=\phi_{AB}\epsilon_{A'B'}+\bar{\phi}_{A'B'}\epsilon_{AB}.
\]
The Maxwell equations are equivalent to the \emph{spin-1 zero-rest mass field equations}:
\[
\nabla^{AA'}\phi_{AB}=0.
\]
If the conformal weight of $\phi_{AB}$ is chosen properly, the vacuum
Maxwell equations are conformally invariant. If $\tilde{\phi}_{AB}$
denotes the \emph{physical} Maxwell spinor, then the Maxwell spinor in
the conformally rescaled (unphysical) spacetime is given by
\[
\phi_{AB}=\Theta^{-1}\tilde{\phi}_{AB}.
\] 

Due to the totally symmetric character of $\phi_{AB}$, one can write
\[
\phi_{AB}=\phi_0\epsilon^0_{AB} +\phi_1\epsilon^1_{AB}+\phi_2 \epsilon^2_{AB},
\]
where the totally symmetric spinors $\epsilon^0_{AB}$,
$\epsilon^1_{AB}$ and $\epsilon^2_{AB}$ are defined in appendix
\ref{appendix:spinors}. In the F-gauge, and using a space-spinor
decomposition, the Maxwell equations can be shown to imply the
following system of 4 equations:
\begin{subequations}
\begin{eqnarray}
&& (\sqrt{2}-2c^0_{01})\partial_\tau \phi_0 +2c^0_{00}\partial_\tau \phi_1 -2c^\alpha_{01}\partial_\alpha\phi_0+2c^\alpha_{00}\partial_\alpha\phi_1 \nonumber \\
&&\hspace{3cm}=(2\Gamma_{0011}-4\Gamma_{1010})\phi_0+4\Gamma_{1000}\phi_1-2\Gamma_{0000}\phi_2, \label{maxwell_1}\\
&& (\sqrt{2}-2c^0_{01})\partial_\tau \phi_1 +2c^0_{00}\partial_\tau \phi_2 -2c^\alpha_{01}\partial_\alpha\phi_1+2c^\alpha_{00}\partial_\alpha\phi_2 \nonumber \\
&&\hspace{3cm}=-2\Gamma_{0111}\phi_0+4\Gamma_{0011}\phi_1-(3\Gamma_{0001}+2\Gamma_{0100})\phi_2, \label{maxwell_2}\\
&& (\sqrt{2}+2c^0_{01})\partial_\tau \phi_1 -2c^0_{11}\partial_\tau\phi_0 + 2c^\alpha_{01}\partial_\alpha \phi_1-2c^\alpha_{11}\partial_\alpha\phi_0 \nonumber \\
&&\hspace{3cm}=(2\Gamma_{0111}-4\Gamma_{1110})\phi_0+4\Gamma_{1100}\phi_1+(2\Gamma_{0100}-3\Gamma_{0001})\phi_2, \label{maxwell_3}\\
&& (\sqrt{2}+2c^0_{01})\partial_\tau \phi_2 -2c^0_{11}\partial_\tau\phi_1 + 2c^\alpha_{01}\partial_\alpha \phi_2-2c^\alpha_{11}\partial_\alpha\phi_1 \nonumber \\
&&\hspace{3cm}=-2\Gamma_{1111}\phi_0+4\Gamma_{0111}\phi_1+(2\Gamma_{1100}-4\Gamma_{0101})\phi_2. \label{maxwell_4}
\end{eqnarray}
\end{subequations}
This system of equations resembles in its form the Maxwell equations
in the Newman-Penrose formalism ---see for example \cite{Ste91}.
Taking linear combinations of equations (\ref{maxwell_2}) and
(\ref{maxwell_3}) one recovers the system of \emph{propagation
  equations} and the \emph{constraint equation} given in
\cite{Val07b}. Particularising to the case of a Schwarzschild
background one obtains
\begin{subequations}
\begin{eqnarray*}
&& (1-c^0)\partial_\tau \phi_0 - c^1\partial_\rho \phi_0 + c^+X_+\phi_1-\Gamma_0\phi_0=0, \\
&& (1-c^0)\partial_\tau \phi_1 - c^1\partial_\rho \phi_1 + c^+X_+\phi_2-\Gamma_1\phi_1=0, \\
&& (1+c^0) \partial_\tau \phi_1 + c^1\partial_\rho \phi_1 + c^-X_-\phi_0-\Xi_1 \phi_1=0, \\ 
&& (1+c^0) \partial_\tau \phi_2 + c^1\partial_\rho \phi_2 + c^-X_-\phi_1-\Xi_2 \phi_2=0.
\end{eqnarray*}
\end{subequations}
where $c^0$, $c^1$, $c^\pm$ are the analytic functions in the
coordinates $(\tau,\rho)$ solving the system
(\ref{system:Schwarzschild}) discussed in section \ref{F_gauge}. The
functions $\Gamma_0$, $\Gamma_1$, $\Xi_1$ and $\Xi_2$ are linear
combinations of the analytic \emph{components of the connection} $f$, $\xi$,
$\chi_2$ and $\chi_h$. More precisely,
\begin{eqnarray*}
&& \Gamma_0=2\Gamma_{0011}-4\Gamma_{1010}=2\xi +\frac{\sqrt{2}}{6}\chi_2 -2\sqrt{2}\chi_h -\sqrt{2}f, \\
&& \Gamma_1=4\Gamma_{0011}= 3\xi -\frac{5\sqrt{2}}{12}\chi_2 -\frac{5\sqrt{2}}{2}\chi_h, \\
&& \Xi_1=-4\Gamma_{1100}=3\xi-\frac{\sqrt{2}}{4}\chi_2 +\frac{3\sqrt{2}}{2}\chi_h, \\
&& \Xi_2= 2\Gamma_{1100}-4\Gamma_{0101}= -2\xi +\frac{\sqrt{2}}{6}\chi_2 -2\sqrt{2}\chi_h +\sqrt{2}f.
\end{eqnarray*}
The precise form of these functions will not be required in our
discussion ---a list of the leading terms of their asymptotic
expansions in the gauge for which $\kappa=\rho$ can be found in the
appendix of \cite{Val07b}. In the sequel, it will be convenient to
isolate the leading terms of these functions. We write:
\begin{subequations}
\begin{eqnarray*}
&& c^0=-\tau- a, \quad c^1=\rho + b, \\
&& c^+= 1+ c, \quad c^-= 1+ c, \\
&& \Gamma_0 = -1- f_0, \quad \Gamma_1 = -f_1, \\
&& \Xi_1= -g_0, \quad \Xi_2= 1 -g_1, 
\end{eqnarray*}
\end{subequations}
where $a$, $b$, $c$, $f_0$, $f_1$, $g_0$, $g_1$ are analytic functions
of $\tau$, $\rho$ independent of the ``angular coordinates''
$\varsigma\in SU(2)$. One has that
\begin{eqnarray*}
&& a=\mathcal{O}(\rho), \quad b=\mathcal{O}(\rho^2), \quad c=\mathcal{O}(\rho), \\
&& f_0=\mathcal{O}(\rho), \quad f_1=\mathcal{O}(\rho), \quad g_0=\mathcal{O}(\rho), \quad g_1=\mathcal{O}(\rho). 
\end{eqnarray*}
Further, they all vanish if $m=0$ where $m$ is the mass of the Schwarzschild spacetime. Due to their monopolar nature
\begin{eqnarray*}
&&X_\pm a=X_\pm b =X_\pm c =X_\pm f_0 =X_\pm f_1=X_\pm g_0=X_\pm g_1=0,  \\
&&X a=X b =X c =X f_0 =X f_1 =X g_0 =X g_1=0, 
\end{eqnarray*}
where $X_+$, $X_-$ and $X$ are the differential operators on $SU(2)$
discussed in section \ref{F_gauge}. Using this notation, the equations
take the form
\begin{subequations}
\begin{eqnarray}
&& A_0\equiv (1+\tau +a )\partial_\tau \phi_0 - (\rho+b)\partial_\rho \phi_0 + (1+c)X_+\phi_1+(1+f_0)\phi_0=0, \label{A0}\\
&& A_1\equiv (1+\tau +a )\partial_\tau \phi_1 - (\rho+b)\partial_\rho \phi_1 + (1+c)X_+\phi_2 + f_1\phi_1=0, \label{A1}\\
&& B_1\equiv (1-\tau - a) \partial_\tau \phi_1 + (\rho+b)\partial_\rho \phi_1 +(1+c)X_-\phi_0 +g_0 \phi_1=0, \label{B1}\\ 
&&  B_2\equiv (1-\tau-a) \partial_\tau \phi_2 + (\rho+b)\partial_\rho \phi_2 + (1+c)X_-\phi_1-(1-g_1)\phi_2=0, \label{B2}
\end{eqnarray}
\end{subequations}
which is the form that will be used in the rest of the article. 

\subsubsection*{On the characteristics of the Maxwell equations in the F-gauge}
In the following, certain questions concerning the characteristics of
the equations (\ref{A0})-(\ref{B2}). From general theory, these have
to coincide with null hypersurfaces in $\mathcal{M}_{a_*,\omega}$ and
can be grouped in \emph{outgoing} and \emph{incoming} according to
whether they intersect future or past null infinity. Let
$(\check{\tau}(s),\check{\rho}(s))$ be the solutions to the system of
ordinary differential equations
\begin{eqnarray*}
&& \frac{\mbox{d}\check{\tau}(s)}{\mbox{d}s}=\bigg(1+\check{\tau}(s)+a(\check{\tau}(s),\check{\rho}(s))\bigg), \quad \check{\tau}(0)=0, \\
&& \frac{\mbox{d}\check{\rho}(s)}{\mbox{d}s}=-\bigg(\check{\rho}(s)+b(\check{\tau}(s),\check{\rho}(s)) \bigg), \quad \check{\rho}(0)=\rho_\bullet,
\end{eqnarray*}
for $0\leq\rho_\bullet < a_*$. Similarly, let $(\hat{\tau}(s),\hat{\rho}(s))$ be the solutions of
\begin{eqnarray*}
&& \frac{\mbox{d}\hat{\tau}(s)}{\mbox{d}s}=\bigg(1-\hat{\tau}(s)-a(\hat{\tau}(s),\hat{\rho}(s))\bigg), \quad \hat{\tau}(0)=0, \\
&& \frac{\mbox{d}\hat{\rho}(s)}{\mbox{d}s}=\bigg(\hat{\rho}(s)+b(\hat{\tau}(s),\hat{\rho}(s)) \bigg), \quad \hat{\rho}(0)=\rho_\bullet.
\end{eqnarray*}
In the case of Minkowski spacetime, the solutions to the above
equations as given ---after a change of parameter--- by
\begin{eqnarray*}
&& \check{\tau}(s')=\frac{s'}{1-s'}, \quad \check{\rho}(s')=\rho_\bullet(1-s'), \\
&& \hat{\tau}(s')=\frac{s'}{1+s'}, \quad \hat{\rho}(s')=\rho_\bullet(1+s'),
\end{eqnarray*}
for $s'\in[-1/2,1/2]$. As in the case of the Minkowski spacetime, it can be
shown that for $(\check{\tau}(s),\check{\rho}(s))$ and $\rho_\bullet\neq 0$,
there is a $s_{\scri^+}\in \Real$ such that $\check{\tau}(s_{\scri^+})=1$ and
$\check{\rho}(s_{\scri^+})\neq 0$. On the other hand one can show that
$\hat{\tau}(s) \neq 1$ for all $s$ ---in fact, $\hat{\tau}(s)\rightarrow 1$ as
$s\rightarrow \infty$. Analogous statements can be made for past null
infinity. In accordance with the previous discussion
\begin{eqnarray*}
&& \check{\mathcal{B}}_{\rho_\bullet} =\bigg\{ (\tau,\rho,\varsigma)\in \mathcal{M}_{a,\omega} \;\bigg| \; \tau=\check{\tau}(s), \; \rho=\check{\rho}(s), \; s\in(s_-,s_+)  \bigg\}, \\
&& \hat{\mathcal{B}}_{\rho_\bullet} =\bigg\{ (\tau,\rho,\varsigma)\in \mathcal{M}_{a,\omega} \;\bigg| \; \tau=\hat{\tau}(s), \; \rho=\hat{\rho}(s), \; s\in(s_-,s_+)  \bigg\},
\end{eqnarray*}
correspond, respectively, to \emph{outgoing} and \emph{incoming}
characteristics of (\ref{A0})-(\ref{B2}). As discussed in \cite{Fri98a} as
$\rho_0\rightarrow 0$ the sets $\check{\mathcal{B}}_{\rho_\bullet}$ and
$\hat{\mathcal{B}}_{\rho_\bullet}$ approach $\mathcal{I}\cup \scri^-$ and
$\mathcal{I} \cup \scri^+$, respectively, in a non-uniform manner. In this
sense, the set $\mathcal{I}$ can be regarded as a limit set of both incoming
and outgoing geodesics ---a \emph{total characteristic}.

\subsubsection*{On the degeneracy of the propagation equations at the conformal boundary}
As discussed in \cite{Fri98a,Fri03b,Val07b} a structural property of
propagation equations derived from a covariant equation with principal
part of the form $\nabla^{AA'}\phi_{A\cdots P}$ ---as in the case of
the propagation equations (\ref{A0})-(\ref{B2})--- is the degeneracy
of subsets of them at the critical sets $\mathcal{I}^\pm$. This can be
readily seen in equations (\ref{A0})-(\ref{B2}) by noting that
$a|_{\mathcal{I}^\pm}=0$ and hence
\[
(1+\tau+a)\big|_{\mathcal{I}^+}=0, \quad (1-\tau-a)\big|_{\mathcal{I}^-}=0.
\]
This state of affairs can be ``worsened'' by some choices of conformal
gauge. In particular if the function $\kappa$ in the conformal factor
(\ref{cf:Theta}) is chosen such that $\kappa=\omega$ ---as it is done
in the present calculations--- then using (\ref{c0_scri_1}) and
(\ref{c0_scri_2}) one has that
\[
(1+\tau+a)\big|_{\scri^+}=0, \quad (1-\tau-a)\big|_{\scri^-}=0.
\]
It turns out that this ``worsening'' of the degeneracy of the
propagation equations at the conformal boundary does not complicates
the analysis any further, and actually makes some of the calculations
easier.

\subsubsection*{Asymptotic expansions}
As discussed in \cite{Val07b} the structural properties of the
equations (\ref{A0})-(\ref{B2}) allow to calculate asymptotic
expansions of the components $\phi_0$, $\phi_1$ and $\phi_2$ of the
form
\begin{equation}
\phi_k \sim \sum_{l\geq |1-k|}\frac{1}{l!} \phi_k^{(l)} \rho^l, 
\quad \phi^{(l)}_k= \partial^l_\rho \phi_k\big|_{\rho=0}, \label{asymptotic:expansion}
\end{equation}
for $k=0$, $1$, $2$. The coefficients in the expansion
(\ref{asymptotic:expansion}) are determined by exploiting the fact
that the cylinder at infinity $\mathcal{I}$ is a total characteristic
of equations (\ref{A0})-(\ref{B2}): the equations reduce to a system
of interior equations when evaluated on $\mathcal{I}$. To exploit this feature
it shall be assumed that the initial data for the equations
(\ref{A0})-(\ref{B2}) on $\mathcal{C}_{a_*}$ are of the form
\begin{equation}
\phi_j = \sum^\infty_{l= |1-j|} \sum_{q=|1-j|}^l \sum_{k=0}^{2q}  \frac{1}{l!}
w_{j,l;q,k} \TT{2q}{k}{q+j-1}\rho^l, \label{initial:data}
\end{equation}
for $j=0,1,2$ with $w_{j,l;q,k}\in \Complex$. The initial data is subject to
the constraint
\[
\rho \partial_\rho \phi_1 + \frac{1}{2}X_-\phi_0 -\frac{1}{2}X_+\phi_2=0,
\]
so that all the coefficients in the expansions of $\phi_1$ ---except for
$w_{1,0;0,0}$, the electric charge--- are determined from those of $\phi_0$
and $\phi_2$. 

Using initial data of the form (\ref{initial:data}) it is possible to calculate
the asymptotic expansions (\ref{asymptotic:expansion}) ---that is, the
$(\tau,\varsigma)$-dependent coefficients $\phi_k^{(l)}$, $\varsigma\in SU(2)$--- to any
desired order ---the only limitation being the computational
complexities. This procedure is a way of
unfolding the evolution process so that it can be analysed in detail
and to any order. In particular, the expansions allow to relate
properties of the initial data with behaviour at null infinity. 

The properties of the coefficients $\phi^{(l)}_k$ and the occurrence
of logarithmic singularities in them at the critical sets
$\mathcal{I}^\pm$, has been the topic of \cite{Val07b}. Here, we start
to analyse the way in which the expansions
(\ref{asymptotic:expansion}) are related to actual solutions of
(\ref{A0})-(\ref{B2}). To this end it will be assumed that given an
integer $p>0$ the solutions of the (\ref{A0})-(\ref{B2}) are of the
form
\begin{equation}
\phi_k = \sum^{p-1}_{l=|1-k|}\frac{1}{l!} \phi_k^{(l)} \rho^l + R_p(\phi_k), 
\quad \phi^{(l)}_k= \partial^l_\rho \phi_k\big|_{\rho=0}, \label{Ansatz}
\end{equation} 
with the residue of order $p$ of $\phi_k$, $R_p(\phi_k)$, given by
\[
R_p(\phi_k)= J^p(\partial^p_\rho\phi_k),
\]
where $J$ denotes the operator $f \mapsto J(f)$ such that
\[
J(f)(\rho)=\int_0^\rho f(s) \mbox{d}s.
\]

The objective of this article is to obtain estimates on the solutions
to (\ref{A0})-(\ref{B2}) by exploiting the Ansatz (\ref{Ansatz}). As
the coefficients $\phi_k^{(l)}$, for $0\leq l \leq p$ and $k=0$, $1$,
$2$ are in principle known explicitly ---and hence also their
regularity---, the latter implies obtaining estimates on
$\partial^p_\rho \phi_k$.

\emph{In what follows it will be assumed that the coefficients
  $\phi_k^{(l)}$ solution of the Maxwell transport equations on
  $\mathcal{I}$ are as smooth as necessary for our calculations to
  make sense.} This assumption implies, in turn, a restriction on the
class of initial data on $\mathcal{C}_{a_*}$ to be considered ---that
is, the coefficients in the expansions (\ref{initial:data}) for
$j=0,2$ should satisfy some regularity conditions. The discussion in
\cite{Val07b} shows how this can be done.

\section{Construction of estimates in Minkowski spacetime} \label{section:Minkowski}

As pointed out in \cite{Fri03b} for the case of the spin-2 field on
flat spacetime, the degeneracy of the propagation equations at the
critical sets $\mathcal{I}^\pm$ has as consequence that estimates
obtained using the standard argument for symmetric hyperbolic
equations ---see e.g. \cite{Joh91,Tay96}--- are not bounded at
$\mathcal{I}^\pm$, and depending on the precise details of the
conformal gauge, possibly also not bounded at $\scri^\pm$. In
\cite{Fri03b} it was possible to overcome this difficulty by noticing
that although it is not possible to obtain directly estimates of
$\phi_k$, it is possible to obtain estimates for $\partial^p_\rho
\phi_k$. This information can be used to control the residue in the
expansion (\ref{Ansatz}).

The methods in \cite{Fri03b} can be readily transcribed for the case
of the Maxwell field. For completeness we present a summary of the
arguments.  It will serve to motivate the discussion for the curved
spacetime case and also to highlight the new complications that arise.

In the case of the Minkowski spacetime and with the choice of
conformal gauge given by $\kappa=\omega=\rho$, equations
(\ref{A0})-(\ref{B2}) take the form
\begin{subequations}
\begin{eqnarray}
&& A_k \equiv (1+\tau) \partial_\tau \phi_k -\rho \partial_\rho \phi_k + X_+ \phi_{k+1} +(1-k)\phi_k, \label{M:M1}\\
&& B_k \equiv (1-\tau) \partial_\tau \phi_{k+1} +\rho \partial_\rho \phi_{k+1} + X_-\phi_k -k \phi_{k+1}, \label{M:M2}
\end{eqnarray}
\end{subequations}
with $k=0$, $1$. As it is usual in the construction of estimates, it
is assumed that one has a solution $\phi_k$, $k=0,1,2$ of the required
smoothness. Consider
\begin{equation}
D^{q,p,\alpha} \overline{\phi}_k D^{q,p,\alpha}A_k + D^{q,p,\alpha} \phi_k D^{q,p,\alpha} \overline{A}_k + D^{q,p,\alpha} \overline{\phi}_{k+1} D^{q,p,\alpha}B_k + D^{q,p,\alpha} \phi_{k+1} D^{q,p,\alpha} \overline{B}_k=0,  \label{Minkowski_integrand_1}
\end{equation}
for $k=0,1$. The notation
\[
D^{q,p,\alpha} =\partial^q_\tau \partial^p_\rho Z^\alpha =\partial^q_\tau
\partial^p_\rho Z_1^{\alpha_1}Z_2^{\alpha_2}Z_3^{\alpha_3}, \quad
\alpha=(\alpha_1,\alpha_2,\alpha_3), \quad |\alpha|=\alpha_1+\alpha_2+\alpha_3, 
\]
has been introduced and will be used throughout the rest of the
article. The operators $Z_1$, $Z_2$ and $Z_3$ denote the differential
operators over $SU(2)$ discussed in appendix \ref{appendix:SU(2)}. These
operators can be written as linear combinations of the operators $X_\pm$ and
$X$ discussed in section (\ref{F_gauge}). 

The expression (\ref{Minkowski_integrand_1}) can be rewritten as
\begin{eqnarray}
&& 0= \partial_\tau \left( (1+\tau) |D^{q,p,\alpha}\phi_k|^2 + (1-\tau) |D^{q,p,\alpha}\phi_{k+1}|^2 \right) + \partial_\rho  \left( \rho |D^{q,p,\alpha}\phi_{k+1}|^2 -\rho |D^{q,p,\alpha}\phi_k |^2 \right) \nonumber \\
&& \hspace{1cm} +\left( D^{q,p,\alpha} \overline{\phi}_k D^{q,p,\alpha} X_+\phi_{k+1} + D^{q,p,\alpha}\phi_{k+1} D^{q,p,\alpha}X_+\overline{\phi}_k \right)  \nonumber \\
&& \hspace{1cm} + \left(D^{q,p,\alpha}\phi_k D^{q,p,\alpha}X_-\overline{\phi}_{k+1} + D^{q,p,\alpha}\overline{\phi}_{k+1} D^{q,p,\alpha} X_- \phi_k\right) \nonumber \\
&& \hspace{1cm} -2(p-q+k-1)|D^{q,p,\alpha} \phi_k|^2 + 2(p-q-k) |D^{q,p,\alpha} \phi_{k+1}|^2. \label{estimate:integrand}
\end{eqnarray} 
For $t\in[0,1]$ and $0<\rho_*<a_*$, introduce the following subsets of $\mathcal{M}_{{a_*},\omega}\approx [-1,1]\times [0,\infty)\times SU(2)$:
\begin{subequations}
\begin{eqnarray}
&& \mathcal{N}_t =\bigg\{ (\tau,\rho,\varsigma)\in \mathcal{M}_{{a_*},\omega} \;\bigg|\;  0 \leq \tau \leq t, \; 0\leq \rho \leq \frac{\rho_*}{1+\tau} \bigg\}, \label{M:domain1}\\
&& \mathcal{B}_t =\bigg\{  (\tau,\rho,\varsigma)\in \mathcal{M}_{{a_*},\omega} \;\bigg|\; 0 \leq \tau \leq t, \; \rho =\frac{\rho_*}{1+\tau} \bigg\}, \label{M:domain2}\\
&& \mathcal{S}_t =\bigg\{ (\tau,\rho,\varsigma)\in \mathcal{M}_{{a_*},\omega} \;\bigg|\; \tau= t, \; 0\leq \rho \leq \frac{\rho_*}{1+\tau} \bigg\},  \label{M:domain3}\\
&& \mathcal{I}_t =\bigg\{ (\tau,\rho,\varsigma)\in \mathcal{M}_{{a_*},\omega} \;\bigg|\; 0 \leq \tau \leq t, \; \rho=0  \bigg \}. \label{M:domain4}
\end{eqnarray}
\end{subequations}
Hence $\mathcal{N}_t$ is the domain of influence of $\mathcal{S}_0\subset
\mathcal{C}_{a_*}$. Let $\mbox{d}\mu$ denote the Haar measure over $SU(2)$.
Integrating expression (\ref{estimate:integrand}) over $\mathcal{N}_t$ and
noting that because of Gauss' theorem
\begin{eqnarray*}
&& \int_{\mathcal{N}_t} \left(\partial_\tau \bigg( (1+\tau) |D^{q,p,\alpha}\phi_k|^2 + (1-\tau) |D^{q,p,\alpha}\phi_{k+1}|^2 \bigg) + \partial_\rho  \bigg( \rho |D^{q,p,\alpha}\phi_{k+1}|^2 -\rho |D^{q,p,\alpha}\phi_k |^2 \bigg) \right)\mbox{d}\tau \mbox{d}\rho \mbox{d}\mu \nonumber \\
&& \hspace{1cm} = \int_{\mathcal{S}_t} \bigg( (1+t)|D^{q,p,\alpha}\phi_k|^2 + (1-t) |D^{q,p,\alpha}\phi_{k+1}|^2 \bigg) \mbox{d}\rho \mbox{d}\mu - \int_{\mathcal{S}_0} \bigg( |D^{q,p,\alpha} \phi_k|^2 + |D^{q,p,\alpha}\phi_{k+1}|^2 \bigg)\mbox{d}\rho \mbox{d}\mu \nonumber \\
&& \hspace{1cm} \phantom{=} +\int_{\mathcal{B}_t} \bigg(\big( (1+\tau) n_\tau -\rho n_\rho\big) |D^{q,p,\alpha} \phi_k|^2 +\big((1-\tau)n_\tau+\rho n_\rho\big)|D^{q,p,\alpha}\phi_{k+1}|^2\bigg) \mbox{d}n \mbox{d}\mu \nonumber \\
&&  \hspace{1cm} \phantom{=} -\int_{\mathcal{I}_t} \rho\big( |D^{q,p,\alpha} \phi_{k+1}|^2 -|D^{q,p,\alpha}\phi_k|^2 \big) \mbox{d}\tau \mbox{d}\mu
\end{eqnarray*}
one obtains
\begin{eqnarray}
&& \int_{\mathcal{S}_t} \bigg( (1+t)|D^{q,p,\alpha}\phi_k|^2 + (1-t) |D^{q,p,\alpha}\phi_{k+1}|^2 \bigg) \mbox{d}\rho \mbox{d}\mu \nonumber \\
&& \hspace{1cm} +\int_{\mathcal{N}_t} \left( D^{q,p,\alpha} \overline{\phi}_k D^{q,p,\alpha} X_+\phi_{k+1} + D^{q,p,\alpha} \phi_{k+1} D^{q,p,\alpha}X_+\overline{\phi}_k \right) \mbox{d}\tau \mbox{d}\rho \mbox{d}\mu \nonumber \\
&& \hspace{1cm} +\int_{\mathcal{N}_t} \left(D^{q,p,\alpha}\phi_k D^{q,p,\alpha}X_-\overline{\phi}_{k+1} + D^{q,p,\alpha}\overline{\phi}_{k+1} D^{q,p,\alpha} X_- \phi_k\right) \mbox{d}\tau \mbox{d}\rho \mbox{d} \mu \nonumber \\
&& \hspace{1cm} +2(p-q-k) \int_{\mathcal{N}_t} |D^{q,p,\alpha}\phi_{k+1}|^2 \mbox{d}\tau \mbox{d} \rho \mbox{d}\mu \nonumber \\
&& \hspace{2cm} \leq \int_{\mathcal{S}_0} \big( |D^{q,p,\alpha}\phi_k|^2 + |D^{q,p,\alpha}\phi_{k+1}|^2\big) \mbox{d} \rho \mbox{d} \mu + 2(p-q+k-1) \int_{\mathcal{N}_t} |D^{q,p,\alpha} \phi_k |^2 \mbox{d} \tau \mbox{d} \rho \mbox{d} \mu. \nonumber \label{M:inequality1} \\
&& 
\end{eqnarray}
To obtain this last inequality it has been used that
\[
\int_{\mathcal{I}_t} \rho\big( |D^{q,p,\alpha} \phi_{k+1}|^2 -|D^{q,p,\alpha}\phi_k|^2 \big) \mbox{d}\tau \mbox{d}\mu =0,
\]
and that
\[
\int_{\mathcal{B}_t} \bigg(\big( (1+\tau)n_\tau -\rho n_\rho\big) |D^{q,p,\alpha} \phi_k|^2 +\big((1-\tau)n_\tau+\rho n_\rho\big)|D^{q,p,\alpha}\phi_{k+1}|^2\bigg) \mbox{d}n \mbox{d}\mu \geq 0,
\]
with $n_\tau=\nu \rho $ and $n_\rho =\nu(1+\tau)$ for a suitable
normalisation factor $\nu$ ---recall that on the characteristic
$\mathcal{B}_t$ one has $\rho=\rho_*/(1+\tau)$.
 
Now, consider the shifts $q\rightarrow q'$ and $p \rightarrow p+p'$ so
that $ D^{q,p,\alpha} \phi_k \rightarrow D^{q',p+p',\alpha}(\phi_{k})=
D^{q',p'\alpha} (\partial^p_\rho \phi_k)$ and $ D^{q,p,\alpha}
\phi_{k+1} \rightarrow D^{q',p+p',\alpha}(\phi_{k+1})=D^{q',p'\alpha}
(\partial^p_\rho \phi_{k+1})$. Using lemma \ref{lemma:Liegroups} in
appendix \ref{appendix:SU(2)} one has that for $m\in \Natural\cup
\{0\}$,
\begin{eqnarray*}
&&\hspace{-2cm}\sum_{q'+p'+|\alpha|\leq m} \int_{\mathcal{N}_t} \left( D^{q',p',\alpha} (\partial^p_\rho\overline{\phi}_k) D^{q',p',\alpha} X_+(\partial^p_\rho\phi_{k+1}) + D^{q',p',\alpha} (\partial^p_\rho\phi_{k+1}) D^{q',p',\alpha}X_+(\partial^p_\rho\overline{\phi}_k) \right) \mbox{d}\tau \mbox{d}\rho \mbox{d}\mu \nonumber \\ 
&&\hspace{-1cm}=\sum_{q'+p'+|\alpha|\leq m} \int_{\mathcal{N}_t} \left( Z^\alpha (\partial^{q'}_\tau\partial^{p'+p}_\rho\overline{\phi}_k) Z^\alpha (X_+\partial^{q'}_\tau\partial^{p'+p}_\rho\phi_{k+1}) + Z^\alpha (\partial^{q'}_\tau \partial^{p'+p}_\rho \phi_{k+1}) Z^\alpha (X_+\partial^{q'}_\tau\partial^{p'+p}_\rho\overline{\phi}_k) \right) \mbox{d}\tau \mbox{d}\rho \mbox{d}\mu \nonumber  \\
&& \hspace{-1cm} =0,
\end{eqnarray*} 
where $|\alpha|=\alpha_1+\alpha_2+\alpha_3$. Similarly, one has that 
\[
\sum_{q'+p'+|\alpha| \leq m} \int_{\mathcal{N}_t}  \left(D^{q',p',\alpha}(\partial^p_\rho\phi_k) D^{q',p',\alpha}X_-(\partial^p_\rho\overline{\phi}_{k+1}) + D^{q',p',\alpha}(\partial^p_\rho\overline{\phi}_{k+1}) D^{q',p',\alpha} X_-(\partial^p_\rho\phi_k) \right) \mbox{d}\tau \mbox{d}\rho \mbox{d} \mu =0. 
\]
Hence summing the shifted version of inequality (\ref{M:inequality1}) over $q'+p'+\alpha \leq m$, with $m$ a suitable positive integer, one obtains
\begin{eqnarray*}
&& \sum_{q'+p'+|\alpha| \leq m} \int_{\mathcal{S}_t}\bigg( (1+t)|D^{q',p'\alpha}(\partial^p_\rho \phi_k)|^2 + (1-t)|D^{q',p'\alpha}(\partial^p_\rho \phi_{k+1})|^2 \bigg) \mbox{d}\rho \mbox{d}\mu \nonumber \\
&& \hspace{1cm} + 2\sum_{q'+p'+|\alpha|\leq m} (p+p'-q-k) \int_{\mathcal{N}_t} |D^{q',p',\alpha}(\partial^p_{\rho}\phi_{k+1})|^2 \mbox{d} \tau \mbox{d}\rho \mbox{d} \mu \nonumber \\
&& \hspace{2cm} \leq   \sum_{q'+p'+|\alpha| \leq m} \int_{\mathcal{S}_0} \bigg( |D^{q',p',\alpha} (\partial^p_\rho\phi_k)|^2 + |D^{q',p',\alpha} (\partial^p_\rho\phi_{k+1})|^2\bigg)\mbox{d}\rho \mbox{d} \mu \nonumber \\
&& \hspace{3cm} + 2 \sum_{q'+p'+|\alpha| \leq m}(p+p'-q'+k-1) \int_{\mathcal{N}_t} |D^{q',p',\alpha}(\partial^p_\rho\phi_k)|^2 \mbox{d}\rho \mbox{d}\mu. 
\end{eqnarray*}
Now, we note that for $k=0,1$ on the one hand one has  
\[
\sum_{q'+p'+|\alpha|\leq m} (p+p'-q-k) \int_{\mathcal{N}_t} |D^{q',p',\alpha}(\partial^p_{\rho}\phi_{k+1})|^2 \mbox{d} \tau \mbox{d}\rho \mbox{d} \mu  \geq (p-m-1)\sum_{q'+p'+\alpha\leq m}\int_{\mathcal{N}_t} |D^{q',p',\alpha}(\partial^p_{\rho}\phi_{k+1})|^2 \mbox{d} \tau \mbox{d}\rho \mbox{d} \mu,
\] 
and on the other
\[
\sum_{q'+p'+|\alpha|\leq m} (p+p'-q'+k-1)\int_{\mathcal{N}_t} |D^{q',p',\alpha}(\partial^p_\rho\phi_k)|^2 \mbox{d}\tau \mbox{d}\rho \mbox{d}\mu \leq (p+m) \sum_{q'+p'+\alpha\leq m} \int_{\mathcal{N}_t} |D^{q',p',\alpha}(\partial^p_\rho\phi_k)|^2 \mbox{d}\tau \mbox{d}\rho \mbox{d}\mu.  
\]
Hence one obtains the inequality
\begin{eqnarray}
&&\sum_{q'+p'+|\alpha|\leq m} \int_{\mathcal{S}_t} \bigg( (1+t) |D^{q',p',\alpha}(\partial^p_\rho \phi_k)|^2 + (1-t) |D^{q',p',\alpha}(\partial^p_\rho \phi_{k+1})|^2 \bigg) \mbox{d} \rho \mbox{d} \mu \nonumber \\
&&\hspace{1cm}+ 2(p-m-1) \sum_{q'+p'+|\alpha|\leq m} \int_{\mathcal{N}_t} |D^{q',p',\alpha}(\partial^p_\rho \phi_{k+1})|^2 \mbox{d}\tau \mbox{d}\rho \mbox{d}\mu \nonumber \\
&& \hspace{2cm} \leq \sum_{q'+p'+|\alpha|\leq m} \int_{\mathcal{S}_0} \big( |D^{q',p',\alpha}(\partial^p_\rho \phi_k)|^2 + |D^{q',p',\alpha} (\partial^p_\rho \phi_{k+1})|^2 \big) \mbox{d} \rho \mbox{d}\mu \nonumber \\
&& \hspace{4cm}+2(p+m) \sum_{q'+p'+|\alpha|\leq m} \int_{\mathcal{N}_t} |D^{q',p',\alpha}(\partial^p_\rho\phi_k)|^2 \mbox{d}\tau \mbox{d} \rho \mbox{d} \mu. \label{prebasic:inequality}
\end{eqnarray}
The second term on the left hand side of the last inequality can be
disregarded if $p>m+1$ ---the choice
$p=m+1$ is not useful as in the sequel one will need to estimate this terms
as well. Noting that for $t\in[0,1]$
\begin{eqnarray*}
&& \hspace{-1cm}\sum_{q'+p'+|\alpha|\leq m} \int_{\mathcal{S}_t} |D^{q',p',\alpha}(\partial^p_\rho \phi_k)|^2 \mbox{d} \rho \mbox{d} \mu \nonumber \\
&& \hspace{1cm} \leq \sum_{q'+p'+|\alpha|\leq m} \int_{\mathcal{S}_t} \bigg( (1+t) |D^{q',p',\alpha}(\partial^p_\rho \phi_k)|^2 + (1-t) |D^{q',p',\alpha}(\partial^p_\rho \phi_{k+1})|^2 \bigg) \mbox{d} \rho \mbox{d} \mu, 
\end{eqnarray*}
one deduces
\begin{eqnarray}
&& \sum_{q'+p'+|\alpha|\leq m} \int_{\mathcal{S}_t} |D^{q',p',\alpha}(\partial^p_\rho \phi_k)|^2 \mbox{d} \rho \mbox{d} \mu \nonumber \\
&& \hspace{2cm} \leq \sum_{q'+p'+|\alpha|\leq m} \int_{\mathcal{S}_0} \big( |D^{q',p',\alpha}(\partial^p_\rho \phi_k)|^2 + |D^{q',p',\alpha} (\partial^p_\rho \phi_{k+1})|^2 \big) \mbox{d} \rho \mbox{d}\mu \nonumber \\
&& \hspace{4cm} +2(p+m) \sum_{q'+p'+|\alpha|\leq m} \int_{\mathcal{N}_t} |D^{q',p',\alpha}(\partial^p_\rho\phi_k)|^2 \mbox{d}\tau \mbox{d} \rho \mbox{d} \mu, \label{basic:inequality}
\end{eqnarray}
for $k=0,1$. For later use it is noted that
because the first term in inequality (\ref{prebasic:inequality}) is
manifestly positive for $t\in[0,1]$, then if one sets $k=1$ one
obtains
\begin{eqnarray}
&& 2(p-m-1)  \sum_{q'+p'+|\alpha|\leq m} \int_{\mathcal{N}_t} |D^{q',p',\alpha}(\partial^p_\rho \phi_2)|^2 \mbox{d} \tau \mbox{d} \rho \mbox{d}\mu \nonumber \\
&& \hspace{2cm} \leq \sum_{q'+p'+|\alpha|\leq m} \int_{\mathcal{S}_0} \big( |D^{q',p',\alpha}(\partial^p_\rho \phi_1)|^2 + |D^{q',p',\alpha} (\partial^p_\rho \phi_2)|^2 \big) \mbox{d} \rho \mbox{d}\mu \nonumber \\
&& \hspace{4cm} +2(p+m) \sum_{q'+p'+|\alpha|\leq m} \int_{\mathcal{N}_t} |D^{q',p',\alpha}(\partial^p_\rho\phi_1)|^2 \mbox{d}\tau \mbox{d} \rho \mbox{d} \mu. \label{extra:inequality}
\end{eqnarray} 
We now proceed to apply Gronwall's argument to inequaility (\ref{basic:inequality}). One can rewrite the inequality in the form
\begin{equation}
\frac{\mbox{d}y(t)}{\mbox{d}t} \leq f_0 + y(t), \label{Gronwall1}
\end{equation}
where
\begin{subequations}
\begin{eqnarray*}
&& y(t) = \int^t_{0} \left( \int_{\mathcal{S}_t} \sum_{q'+p'+|\alpha|\leq m} |D^{q',p',\alpha} (\partial^p_\rho \phi_k)|^2  \mbox{d} \rho \mbox{d} \mu  \right)\mbox{d} \tau, \\
&& f_0 =  \sum_{q'+p'+|\alpha|\leq m} \int_{\mathcal{S}_0} \big( |D^{q',p',\alpha}(\partial^p_\rho \phi_1)|^2 + |D^{q',p',\alpha} (\partial^p_\rho \phi_2)|^2 \big) \mbox{d} \rho \mbox{d}\mu.
\end{eqnarray*}
\end{subequations}
The integration factor for (\ref{Gronwall1}) is $e^{-2(p+m)t}$, so
that one obtains
\[
\frac{\mbox{d}}{\mbox{d}t}\left(e^{-2(p+m)t}y(t)\right) \leq f_0 e^{-2(p+m)t}.
\]
Hence, after integration, and noting that $y(0)=0$ one obtains
\[
y(t) \leq \frac{\left(e^{2(p+m)t}-1\right)}{2(p+m)}f_0,
\]
or
\begin{eqnarray}
&& \hspace{-1cm} \sum_{q'+p'+|\alpha|\leq m} \int_{\mathcal{N}_t} |D^{q',p',\alpha} (\partial^p_\rho \phi_k)|^2 \mbox{d} \tau \mbox{d} \rho \mbox{d} \mu \nonumber \\
&& \hspace{1cm} \leq \frac{\left( e^{2(p+m)t}-1\right)}{2(p+m)}\sum_{q'+p'+|\alpha|\leq m} \int_{\mathcal{S}_0} \left(|D^{q',p',\alpha} (\partial^p_\rho \phi_k)|^2 + |D^{q',p',\alpha} (\partial^p_\rho \phi_{k+1})|^2 \right) \mbox{d} \rho \mbox{d} \mu, \nonumber \\
&&  \label{Gronwall2}
\end{eqnarray}
for $k=0,1$. An estimate for $D^{q',p',\alpha}(\partial^p_\rho \phi_2)$
can now, in turn, be obtained from inequalities
(\ref{extra:inequality}) and (\ref{Gronwall2}). Accordingly, one obtains
\begin{eqnarray*}
&&\hspace{-1cm} \sum_{q'+p'+|\alpha|\leq m} \int_{\mathcal{N}_t} |D^{q',p',\alpha} (\partial^p_\rho \phi_2)|^2 \mbox{d} \tau \mbox{d} \rho \mbox{d} \mu \\
&& \hspace{1cm} \leq \frac{e^{2(p+m)t}}{2(p-m-1)} \sum_{q'+p'+|\alpha|\leq m} \int_{\mathcal{S}_0} \left( |D^{q',p',\alpha}(\partial^p_\rho \phi_1)|^2 + |D^{q',p',\alpha}(\partial^p_\rho \phi_2)|^2  \right) \mbox{d} \rho \mbox{d} \mu. 
\end{eqnarray*}
Thus, for $k=0,1,2$ and $p>m+1$ one has the estimate
\begin{eqnarray}
&& \int_{\mathcal{N}_t} \left(\sum_{q'+p'+|\alpha|\leq m} |D^{q',p',\alpha} (\partial^p_\rho \phi_k)|^2   \right) \mbox{d} \tau \mbox{d} \rho \mbox{d} \mu \nonumber \\
&& \hspace{2cm} \leq C \sum^2_{k=0} \int_{\mathcal{S}_0} \left( \sum_{q'+p'+|\alpha|\leq m} |D^{q',p',\alpha} (\partial^p_\rho \phi_k)|^2 \right) \mbox{d} \rho \mbox{d} \mu, \label{basic:estimate}
\end{eqnarray}
with $C$ a constant depending on $p$ and $m$ which can be chosen
independently of $t\in[0,1]$.

\subsubsection*{Discussion}
As discussed thoroughly in \cite{Fri03b}, using the Sobolev embedding
theorems one has that given $t\in[0,1]$ and $j=0,$ $1,\ldots$ there
isa continuous embedding $
H^{j+3}(\stackrel{\circ}{\mathcal{N}}_t)\rightarrow
C^{j,\lambda}(\mathcal{N}_t)$ , with $0<\lambda<1$, where $H^{j+3}$
denotes the standard $L^2$-type Sobolev space, and
$\stackrel{\circ}{\mathcal{N}}_t$ denotes the interior of the compact
set $\mathcal{N}_t$. The space $C^{j,\lambda}(\mathcal{N}_t)$ consists
of of functions in $C^j(\stackrel{\circ}{\mathcal{N}}_t)$ which
together with their derivatives up to order $j$ are H\"older
continuous in $\stackrel{\circ}{\mathcal{N}}_t$ and thus, together
with the derivatives, extend to continuous functions on
$\mathcal{N}_t$.

For a $\phi_k$ satisfying the estimate (\ref{basic:estimate}) one has
then that $\partial^p_\rho \phi_k\in C^{j,\lambda}(\mathcal{N}_t)$,
$t\in[0,1]$ if $p\geq j+5$. Consequently, the remainder
$R_p(\phi_k)=J^p(\partial^p_\rho \phi_k)$ in the expansion
(\ref{Ansatz}) is such that $J^p(\partial^p_\rho \phi_k)\in
C^{j,\lambda}(\mathcal{N}_t)$, $t\in[0,1]$. Hence, one can prescribe
the regularity of the remainder in (\ref{Ansatz}) by considering an
expansion with a suitably large order. On the other hand, the
smoothness of the coefficients $\phi^{(l)}_k$, $0\leq l \leq p-1$ is
known explicitly.

Finally, it is noted that if a conformal gauge for which $\kappa\neq
\rho$ in the conformal factor (\ref{cf:Theta}) is used, then it turns
out that the argument for the construction of estimates becomes more
complicated as the coefficients $\partial^{p'}_\rho \phi_k$, $p'<p$
will start appearing in the discussion. In this case, estimates can be
constructed by means of a variant of the construction to be discussed
in the next sections.

\section{Estimates on Schwarzschild spacetime} \label{section:Schwarzschild}
In order to discuss the construction of estimates for equations
(\ref{A0})-(\ref{B2}) for a non-flat background one has
to consider the analogues in the Schwarzschild spacetime of the sets
(\ref{M:domain1})-(\ref{M:domain4}). Given $\rho_*>0$, let
$(\check{\tau}(s),\check{\rho}(s))$ be the solutions of the system
\begin{eqnarray*}
&& \frac{\mbox{d}\check{\tau}(s)}{\mbox{d}s}=\bigg(1+\check{\tau}(s)+a(\check{\tau}(s),\check{\rho}(s))\bigg), \quad \check{\tau}(0)=0, \\
&& \frac{\mbox{d}\check{\rho}(s)}{\mbox{d}s}=-\bigg(\check{\rho}(s)+b(\check{\tau}(s),\check{\rho}(s)) \bigg), \quad \check{\rho}(0)=\rho_*.
\end{eqnarray*}
It can be shown that there is a
$s_{\scri^+}\in \Real^+$such that $\tau(s_{\scri^+})=1$ and
$\rho(s_{\scri^+})\neq 0$ ---that is, the curve intersects future null
infinity. More generally, given $t\in[0,1]$ there is $s_t\in \Real^+$
such that $\check{\tau}(s_t)=t$. Define
\begin{subequations}
\begin{eqnarray*}
&& \mathcal{N}_t =\bigg \{(\tau,\rho,\varsigma)\in \mathcal{M}_{a,\omega} \; \bigg| \; 0\leq \tau \leq t,\; 0\leq \rho \leq \check{\rho}(s), \; s\in[0,s_t] \bigg \}, \\
&& \mathcal{B}_t =\bigg\{  (\tau,\rho,\varsigma)\in \mathcal{M}_{a,\omega} \; \bigg| \; \tau=\check{\tau}(s), \; \rho=\check{\rho}(s), \; s\in[0,s_t]    \bigg\}, \\
&& \mathcal{S}_t =\bigg \{ (\tau,\rho,\varsigma)\in \mathcal{M}_{a,\omega} \; \bigg| \; \tau=t, \; 0\leq \rho \leq \check{\rho}(s), \; s\in[0,s_t] \bigg\},  \\
&& \mathcal{I}_t =\bigg\{ (\tau,\rho,\varsigma)\in \mathcal{M}_{a,\omega} \; \bigg| \;    0\leq \tau \leq t, \; \rho=0 \bigg\},
\end{eqnarray*}
\end{subequations}
for $t\in[0,1]$. Again $\mathcal{N}_t$ is the domain of influence of
$\mathcal{S}_0\subset \mathcal{C}_{a_*}$. 

Given $(p,q,\alpha)$ such that  $q+p+|\alpha|=m$ for a given  $m\in \Natural\cup\{0\}$  a lengthy but straightforward calculation gives
\begin{eqnarray}
&& \hspace{-2cm}D^{q,p,\alpha}A_k =(1+\tau+a) \partial_\tau D^{q,p,\alpha}\phi_k -(\rho+b) \partial_\rho D^{q,p,\alpha} \phi_k + (1+c) D^{q,p,\alpha} X_+ \phi_{k+1} \nonumber \\
&&+ \big((1-k) +f_k -p+q -p D^{0,1,0} b +q D^{1,0,0}a\big) D^{q,p,\alpha}\phi_k \nonumber \\
&& +\bigg( \sum^q_{s=2} \binom{q}{s} D^{s,0,0}a D^{q-s+1,0,0}\phi_k + \sum^q_{s=1} \sum^p_{l=1} \binom{q}{s} \binom{p}{l}D^{s,l,0}a D^{q-s,p-l,\alpha}\phi_k\bigg) \nonumber \\
&& -\bigg( \sum^q_{s=1} \binom{q}{s} D^{s,0,0}b D^{q-s,1,\alpha}\phi_k + \sum^q_{s=0} \sum^p_{l=2} \binom{q}{s} \binom{p}{l}D^{s,l,0}b D^{q-s,p-l+1,\alpha}\phi_k\bigg) \nonumber \\
&& +\bigg( \sum^q_{s=1} \binom{q}{s} D^{s,0,0}c D^{q-s,p,\alpha}X_+\phi_{k+1} + \sum^q_{s=0} \sum^p_{l=1} \binom{q}{s} \binom{p}{l}D^{s,l,0}c D^{q-s,p-l,\alpha}X_+\phi_{k+1}\bigg)  \nonumber \\
&& +\bigg( \sum^q_{s=2} \binom{q}{s} D^{s,0,0}f_k D^{q-s,0,\alpha}\phi_k -p\sum_{s=1}^q\binom{q}{s}D^{s,1,0}f_kD^{q-s,0,\alpha}\phi_k \bigg) \nonumber \\
&& + \sum^q_{s=0} \sum^p_{l=1} \binom{q}{s} \binom{p}{l}D^{s,l,0}f_k D^{q-s,p-l,\alpha}\phi_k=0, \label{DA}
\end{eqnarray}
and
\begin{eqnarray}
&& \hspace{-2cm}D^{q,p,\alpha}B_k =(1-\tau-a) \partial_\tau D^{q,p,\alpha}\phi_{k+1} +(\rho+b) \partial_\rho D^{q,p,\alpha} \phi_{k+1} + (1+c) D^{q,p,\alpha} X_- \phi_{k} \nonumber \\
&&+ \big(-k +g_k +p-q +p D^{0,1,0} b -q D^{1,0,0}a\big) D^{q,p,\alpha}\phi_{k+1} \nonumber \\
&& -\bigg( \sum^q_{s=2} \binom{q}{s} D^{s,0,0}a D^{q-s+1,0,0}\phi_{k+1} + \sum^q_{s=1} \sum^p_{l=1} \binom{q}{s} \binom{p}{l}D^{s,l,0}a D^{q-s,p-l,\alpha}\phi_{k+1}\bigg) \nonumber \\
&& +\bigg( \sum^q_{s=1} \binom{q}{s} D^{s,0,0}b D^{q-s,1,\alpha}\phi_{k+1} + \sum^q_{s=0} \sum^p_{l=2} \binom{q}{s} \binom{p}{l}D^{s,l,0}b D^{q-s,p-l+1,\alpha}\phi_{k+1}\bigg) \nonumber \\
&& +\bigg( \sum^q_{s=1} \binom{q}{s} D^{s,0,0}c D^{q-s,p,\alpha}X_+\phi_{k+1} + \sum^q_{s=0} \sum^p_{l=1} \binom{q}{s} \binom{p}{l}D^{s,l,0}c D^{q-s,p-l,\alpha}X_-\phi_{k}\bigg) \nonumber \\
&& +\bigg( \sum^q_{s=2} \binom{q}{s} D^{s,0,0}g_k D^{q-s,0,\alpha}\phi_{k+1} +p\sum_{s=1}^q\binom{q}{s}D^{s,1,0}g_kD^{q-s,0,\alpha}\phi_{k+1} \bigg) \nonumber \\
&& + \sum^q_{s=0} \sum^p_{l=1} \binom{q}{s} \binom{p}{l}D^{s,l,0}g_{k} D^{q-s,p-l,\alpha}\phi_{k+1}=0, \label{DB}
\end{eqnarray}
for $k=0,1$. It is important to note in these expressions the presence
of terms of the form $D^{q,p,\alpha}\phi_k$, $k=0,1,2$ with
$q+p+|\alpha|\leq m$. These impede the straight-forward application of
the methods discussed in section \ref{section:Minkowski}. However, it
turns out that it is possible to construct estimates like those in
\ref{section:Minkowski} by means of an induction argument.

For later reference it will be convenient to write 
\begin{subequations}
\begin{eqnarray}
&&\hspace{-2cm} H^{(q,j,\alpha)}_a= \sum^q_{s=2} \binom{q}{s} D^{s,0,0}a D^{q-s+1,0,\alpha}\phi_k + \sum^q_{s=1} \sum^{p+j}_{l=1} \binom{q}{s} \binom{p+j}{l}D^{s,l,0}a D^{q-s,p+j-l,\alpha}\phi_k,\label{H_a}\\
&&\hspace{-2cm} H^{(q,j,\alpha)}_b= -\sum^q_{s=1} \binom{q}{s} D^{s,0,0}b D^{q-s,1,\alpha}\phi_k - \sum^q_{s=0} \sum^{p+j}_{l=2} \binom{q}{s} \binom{p+j}{l}D^{s,l,0}b D^{q-s,p+j-l+1,\alpha}\phi_k,\\
&&\hspace{-2cm} H^{(q,j,\alpha)}_c= \sum^q_{s=1} \binom{q}{s} D^{s,0,0}c D^{q-s,p,\alpha}X_+\phi_{k+1} + \sum^q_{s=0} \sum^{p+j}_{l=1} \binom{q}{s} \binom{p+j}{l}D^{s,l,0}c D^{q-s,p+j-l,\alpha}X_-\phi_{k},\\
&&\hspace{-2cm} H^{(q,j,\alpha)}_{f_k}=\sum^q_{s=2} \binom{q}{s} D^{s,0,0}f_k
D^{q-s,0,\alpha}\phi_k
-p\sum_{s=1}^q\binom{q}{s}D^{s,1,0}f_kD^{q-s,0,\alpha}\phi_k \nonumber \\
&& \hspace{3cm}+\sum^q_{s=0} \sum^{p+j}_{l=1} \binom{q}{s} \binom{p+j}{l}D^{s,l,0}f_k D^{q-s,p+j-l,\alpha}\phi_k, \\
&&\hspace{-2cm} K^{(q,j,\alpha)}_a= -\sum^q_{s=2} \binom{q}{s} D^{s,0,0}a D^{q-s+1,0,\alpha}\phi_{k+1} - \sum^q_{s=1} \sum^{p+j}_{l=1} \binom{q}{s} \binom{p+j}{l}D^{s,l,0}a D^{q-s,p+j-l,\alpha}\phi_{k+1}, \\
&&\hspace{-2cm} K^{(q,j,\alpha)}_b= \sum^q_{s=1} \binom{q}{s} D^{s,0,0}b D^{q-s,1,\alpha}\phi_{k+1} + \sum^q_{s=0} \sum^{p+j}_{l=2} \binom{q}{s} \binom{p+j}{l}D^{s,l,0}b D^{q-s,p+j-l+1,\alpha}\phi_{k+1}, \\
&&\hspace{-2cm} K^{(q,j,\alpha)}_c=\sum^q_{s=1} \binom{q}{s} D^{s,0,0}c D^{q-s,p,\alpha}X_+\phi_{k+1} + \sum^q_{s=0} \sum^{p+j}_{l=1} \binom{q}{s} \binom{p+j}{l}D^{s,l,0}c D^{q-s,p+j-l,\alpha}X_+\phi_{k+1}, \\
&&\hspace{-2cm} K^{(q,j,\alpha)}_{g_k}= \sum^q_{s=2} \binom{q}{s} D^{s,0,0}g_k
D^{q-s,0,\alpha}\phi_{k+1}
+p\sum_{s=1}^q\binom{q}{s}D^{s,1,0}g_kD^{q-s,0,\alpha}\phi_{k+1}
\nonumber \\
&& \hspace{3cm}+ \sum^q_{s=0} \sum^{p+j}_{l=1} \binom{q}{s}
\binom{p+j}{l}D^{s,l,0}g_{k} D^{q-s,p+j-l,\alpha}\phi_{k+1}. \label{K_g}
\end{eqnarray}
\end{subequations}

\subsection{Estimates for  $D^{q,0,\alpha} (\partial^p_\rho\phi_k)$ and
  $D^{q,0,\alpha}(\partial^p_\rho\phi_{k+1})$ with $q+|\alpha|\leq m$}
\label{section:base_step} 
First, it is shown how the arguments of section
\ref{section:Minkowski} can be adapted to obtain estimates of $D^{q,0,\alpha}
(\partial^p_\rho\phi_k)$ and $D^{q,0,\alpha}(\partial^p_\rho\phi_{k+1})$ for
$q+|\alpha|\leq m$ and $p$ suitably large ---the latter to be determined during
the argument. Again, as it is customary in the construction of estimates it is
assumed that the relevant objects are as smooth as required. \emph{In
  addition, it will be assumed that the components $\phi_k$ are of the form
(\ref{Ansatz})}.

As in section \ref{section:Minkowski}, the starting point is
\[
D^{q,p,\alpha} \overline{\phi}_k D^{q,p,\alpha}A_k + D^{q,p,\alpha} \phi_k D^{q,p,\alpha} \overline{A}_k + D^{q,p,\alpha} \overline{\phi}_{k+1} D^{q,p,\alpha}B_k + D^{q,p,\alpha} \phi_{k+1} D^{q,p,\alpha} \overline{B}_k=0, 
\]
for $k=0,1$ and $q+|\alpha|\leq m$. A straightforward calculation
using the expressions (\ref{DA}) and (\ref{DB}) shows that the above
can be rewritten as
\begin{eqnarray}
&& 0= \partial_\tau \left( (1+\tau+a) |D^{q,0,\alpha}(\partial^p_\rho\phi_k)|^2 + (1-\tau-a) |D^{q,0,\alpha}(\partial^p_\rho\phi_{k+1})|^2 \right) \nonumber \\
&& \hspace{1cm}+ \partial_\rho  \left( (\rho+b) |D^{q,0,\alpha}(\partial^p_\rho\phi_{k+1})|^2 -(\rho+b) |D^{q,0,\alpha}(\partial^p_\rho\phi_k) |^2 \right) \nonumber \\
&& \hspace{1cm} +(1+c)\left( D^{q,0,\alpha} (\partial^p_\rho\overline{\phi}_k) D^{q,0,\alpha} X_+(\partial^p_\rho\phi_{k+1}) + D^{q,0,\alpha}(\partial^p_\rho\phi_{k+1}) D^{q,0,\alpha}X_+(\partial^p_\rho\overline{\phi}_k) \right)  \nonumber \\
&& \hspace{1cm} + (1+c)\left(D^{q,0,\alpha}(\partial^p_\rho\phi_k) D^{q,p,\alpha}X_-(\partial^p_\rho\overline{\phi}_{k+1}) + D^{q,0,\alpha}(\partial^p_\rho\overline{\phi}_{k+1}) D^{q,0,\alpha} X_- (\partial^p_\rho\phi_k)\right) \nonumber \\
&& \hspace{1cm} -2(p-q+k-1-f_k +pD^{0,p,0}b -qD^{1,0,0}a)|D^{q,0,\alpha}(\partial^p_\rho \phi_k)|^2 \nonumber \\
&& \hspace{1cm}+ 2(p-q-k+g_k +p D^{0,1,0}b -qD^{1,0,0}a) |D^{q,0,\alpha}(\partial^p_\rho \phi_{k+1})|^2. \nonumber \\
&& \hspace{1cm}+ D^{q,0,\alpha}(\partial^p_\rho\overline{\phi}_k) H^{(q,0,\alpha)}_a + D^{q,0,\alpha}(\partial^p_\rho\phi_k) \overline{H}^{(q,0,\alpha)}_a+ D^{q,0,\alpha}(\partial^p_\rho\overline{\phi}_k) H^{(q,0,\alpha)}_b + D^{q,0,\alpha}(\partial^p_\rho\phi_k) \overline{H}^{(q,0,\alpha)}_b \nonumber \\
&& \hspace{1cm}+ D^{q,0,\alpha}(\partial^p_\rho\overline{\phi}_k) H^{(q,0,\alpha)}_c + D^{q,0,\alpha}(\partial^p_\rho\phi_k) \overline{H}^{(q,0,\alpha)}_c+ D^{q,0,\alpha}(\partial^p_\rho\overline{\phi}_k) H^{(q,0,\alpha)}_{f_k} + D^{q,0,\alpha}(\partial^p_\rho\phi_k) \overline{H}^{(q,0,\alpha)}_{f_k} \nonumber\\
&& \hspace{1cm}+ D^{q,0,\alpha}(\partial^p_\rho\overline{\phi}_k) K^{(q,0,\alpha)}_a + D^{q,0,\alpha}(\partial^p_\rho\phi_k) \overline{K}^{(q,0,\alpha)}_a+ D^{q,0,\alpha}(\partial^p_\rho\overline{\phi}_k) K^{(q,0,\alpha)}_b + D^{q,0,\alpha}(\partial^p_\rho\phi_k) \overline{K}^{(q,0,\alpha)}_b \nonumber \\
&& \hspace{1cm}+ D^{q,0,\alpha}(\partial^p_\rho\overline{\phi}_k) K^{(q,0,\alpha)}_c + D^{q,0,\alpha}(\partial^p_\rho\phi_k) \overline{K}^{(q,0,\alpha)}_c+ D^{q,0,\alpha}(\partial^p_\rho\overline{\phi}_k) K^{(q,0,\alpha)}_{f_k} + D^{q,0,\alpha}(\partial^p_\rho\phi_k) \overline{K}^{(q,0,\alpha)}_{g_k}.\nonumber  \\
&& \label{S:integrand}
\end{eqnarray} 
Crucial in this last expression is that the highest $\rho$-derivatives of
$\phi_k$ and $\phi_{k+1}$ in $H_a^{(q,0,\alpha)}$,
$H_b^{(q,0,\alpha)}$,$H_c^{(q,0,\alpha)}$, $H_{f_k}^{(q,0,\alpha)}$,
$K_a^{(q,0,\alpha)}$, $K_b^{(q,0,\alpha)}$, $K_c^{(q,0,\alpha)}$ and
$K_{g_k}^{(q,0,\alpha)}$ ---as given by the expressions
(\ref{H_a})-(\ref{K_g})--- are  of order $p-1$. The use of the Ansatz
(\ref{Ansatz}) will allow to get around the problem of not having estimates for
$\partial^{p'}_\rho\phi_k$ and $\partial^{p'}_\rho\phi_{k+1}$ for $0\leq p'
<p$. Indeed, if
\begin{subequations}
\begin{eqnarray*}
&& \phi_k=\sum_{l=0}^{p-1} \frac{1}{l!}\phi^{(l)}_k \rho^l + J^p(\partial^p_\rho \phi_k), \\
&& \phi_{k+1}=\sum_{l=0}^{p-1} \frac{1}{l!}\phi^{(l)}_{k+1}\rho^l + J^p(\partial^p_\rho \phi_{k+1}),
\end{eqnarray*}
\end{subequations}
then a direct calculation yields
\[
\partial^s_\rho \phi_k = \sum^{p-s-1}_{l=0} \frac{1}{(l-s)!} \phi^{(l)} \rho^{l-s} + J^{p-s}(\partial^p_\rho \phi_k),
\]
for $0\leq s <p$. Thus, for example, one has that
\begin{eqnarray*}
&& H^{(q,0,\alpha)}_a= \sum_{s=2}^q \binom{q}{s}D^{s,0,0}a
D^{q-s+1,0,\alpha}\phi_k + \sum_{s=1}^q \sum_{l=1}^p \binom{q}{s} \binom{p}{l}D^{s,l,0}a
D^{q-s,p-l,\alpha} \phi_k   
\end{eqnarray*}
can be rewritten as
\[
H^{(q,0,\alpha)}_a = F^{(q,0,\alpha)}_a + S^{(q,0,\alpha)}_a
\]
with
\begin{eqnarray*}
&& F^{(q,0,\alpha)}_a= \sum_{s=2}^q \sum_{r=0}^{p-1}\frac{1}{r!} \binom{q}{s}
D^{s,0,0}a D^{q-s+1,0,\alpha} \phi^{(r)}_k \rho ^r \\
&& \hspace{2cm}+ \sum_{s=1}^q
\sum_{l=0}^p\sum_{r=0}^{l-1} \frac{1}{(r-p+l)!} \binom{q}{s} \binom{p}{l}
D^{s,l,0}a D^{q-s,0,\alpha}\phi_k^{(r)}\rho^{r-p+l}, \\
&& S^{(q,0,\alpha)}_a= \sum^q_{s=2} \binom{q}{s} D^{s,0,0} a
J^p(D^{q-s+1,p,\alpha}\phi_k)+ \sum_{s=1}^q
\sum_{l=1}^p\binom{q}{s}\binom{p}{l}D^{s,l,0}a J^l(D^{q-s,p,\alpha}\phi_k).
\end{eqnarray*}
The functions $F^{(q,0,\alpha)}_a$ can be calculated explicitly for given
values of $p$, $q$ and $\alpha$. Moreover, they can be made as smooth as
necessary by an adequate choice of the initial data along the lines of the
discussion in \cite{Val07b}. The terms are homogeneous in
$J^i(D^{j,p,\beta}\phi_k)$, with $1\leq i \leq p$ and $j+|\beta|\leq m$.
Accordingly, given a $\delta>0$ there is a $0<\rho_*<a_*$ such that
$|S^{q,0,\alpha}|< \delta$ on $\mathcal{N}_t$, $t\in[0,1]$. A similar split
can be performed with the other terms in (\ref{S:integrand}) so that one
obtains
\begin{eqnarray*}
&& H^{(q,0,\alpha)}_a= F^{(q,0,\alpha)}_a+S^{(q,0,\alpha)}_a, \quad K^{(q,0,\alpha)}_a=G^{(q,0,\alpha)}_a + R^{(q,0,\alpha)}_a, \\
&& H^{(q,0,\alpha)}_b= F^{(q,0,\alpha)}_b+S^{(q,0,\alpha)}_b, \quad K^{(q,0,\alpha)}_b=G^{(q,0,\alpha)}_b + R^{(q,0,\alpha)}_b, \\
&& H^{(q,0,\alpha)}_c= F^{(q,0,\alpha)}_c+S^{(q,0,\alpha)}_c, \quad K^{(q,0,\alpha)}_c=G^{(q,0,\alpha)}_c + R^{(q,0,\alpha)}_c, \\
&& H^{(q,0,\alpha)}_{f_k}= F^{(q,0,\alpha)}_{f_k}+S^{(q,0,\alpha)}_{f_k}, \quad K^{(q,0,\alpha)}_{g_k}=G^{(q,0,\alpha)}_{g_k} + R^{(q,0,\alpha)}_{g_k},
\end{eqnarray*}
where again $F^{(q,0,\alpha)}_a$, $F^{(q,0,\alpha)}_b$,
$F^{(q,0,\alpha)}_c$, $F^{(q,0,\alpha)}_{f_k}$, $G^{(q,0,\alpha)}_a$,
$G^{(q,0,\alpha)}_b$, $G^{(q,0,\alpha)}_c$ and
$G^{(q,0,\alpha)}_{g_k}$ can be calculated explicitly in terms of the
solutions to the transport equations, and consequently can be made as regular
as necessary by choosing a suitable choice of initial data. On the other hand
$S^{(q,0,\alpha)}_a$,
$S^{(q,0,\alpha)}_b$, $S^{(q,0,\alpha)}_c$, $S^{(q,0,\alpha)}_{f_k}$,
$R^{(q,0,\alpha)}_a$, $R^{(q,0,\alpha)}_b$, $R^{(q,0,\alpha)}_c$ and
$R^{(q,0,\alpha)}_{g_k}$ are homogeneous functions of
$J^i(D^{j,p,\beta}\phi_k)$ and $J^i(D^{j,p,\beta}\phi_{k+1})$ with
$i\geq 1$ and $j+|\beta| \leq m$ and hence their supremum in $\mathcal{N}_t$
can be made suitably small by a convenient choice of $\rho_*$. 

Substitution into (\ref{S:integrand}) gives 
\begin{eqnarray}
&& 0= \partial_\tau \left( (1+\tau+a) |D^{q,0,\alpha}(\partial^p_\rho\phi_k)|^2 + (1-\tau-a) |D^{q,0,\alpha}(\partial^p_\rho\phi_{k+1})|^2 \right) \nonumber \\
&& \hspace{1cm}+ \partial_\rho  \left( (\rho+b) |D^{q,0,\alpha}(\partial^p_\rho\phi_{k+1})|^2 -(\rho+b) |D^{q,0,\alpha}(\partial^p_\rho\phi_k) |^2 \right) \nonumber \\
&& \hspace{1cm} +(1+c)\left( D^{q,0,\alpha} (\partial^p_\rho\overline{\phi}_k) D^{q,0,\alpha} X_+(\partial^p_\rho\phi_{k+1}) + D^{q,0,\alpha}(\partial^p_\rho\phi_{k+1}) D^{q,0,\alpha}X_+(\partial^p_\rho\overline{\phi}_k) \right)  \nonumber \\
&& \hspace{1cm} + (1+c)\left(D^{q,0,\alpha}(\partial^p_\rho\phi_k) D^{q,p,\alpha}X_-(\partial^p_\rho\overline{\phi}_{k+1}) + D^{q,0,\alpha}(\partial^p_\rho\overline{\phi}_{k+1}) D^{q,0,\alpha} X_- (\partial^p_\rho\phi_k)\right) \nonumber \\
&& \hspace{1cm} -2(p-q+k-1-f_k +pD^{0,p,0}b -qD^{1,0,0}a)|D^{q,0,\alpha}(\partial^p_\rho \phi_k)|^2 \nonumber \\
&& \hspace{1cm}+ 2(p-q-k+g_k +p D^{0,1,0}b -qD^{1,0,0}a) |D^{q,0,\alpha}(\partial^p_\rho \phi_{k+1})|^2. \nonumber \\
&& \hspace{1cm}+ D^{q,0,\alpha}(\partial^p_\rho\overline{\phi}_k) \widehat{F}^{(q,0,\alpha)}_k+ D^{q,0,\alpha}(\partial^p_\rho\phi_k) \overline{\widehat{F}}^{(q,0,\alpha)}_k+ D^{q,0,\alpha}(\partial^p_\rho\overline{\phi}_k) \widehat{G}^{(q,0,\alpha)}_k+ D^{q,0,\alpha}(\partial^p_\rho\phi_k) \overline{\widehat{G}}^{(q,0,\alpha)}_k \nonumber \\
&& \hspace{1cm} + D^{q,0,\alpha}(\partial^p_\rho\overline{\phi}_k) \widehat{S}^{(q,0,\alpha)}_k+ D^{q,0,\alpha}(\partial^p_\rho\phi_k) \overline{\widehat{S}}^{(q,0,\alpha)}_k+ D^{q,0,\alpha}(\partial^p_\rho\overline{\phi}_k) \widehat{R}^{(q,0,\alpha)}_k+ D^{q,0,\alpha}(\partial^p_\rho\phi_k) \overline{\widehat{R}}^{(q,0,\alpha)}_k, \nonumber \\ 
&& \label{S:integrand_01}
\end{eqnarray} 
with
\begin{eqnarray*}
&& \widehat{F}^{(q,0,\alpha)}_k = F^{(q,0,\alpha)}_a + F^{(q,0,\alpha)}_b + F^{(q,0,\alpha)}_c + F^{(q,0,\alpha)}_{f_k}, \\
&& \widehat{G}^{(q,0,\alpha)}_k = G^{(q,0,\alpha)}_a + G^{(q,0,\alpha)}_b + G^{(q,0,\alpha)}_c + G^{(q,0,\alpha)}_{g_k}, \\
&& \widehat{S}^{(q,0,\alpha)}_k = S^{(q,0,\alpha)}_a + S^{(q,0,\alpha)}_b + S^{(q,0,\alpha)}_c + S^{(q,0,\alpha)}_{f_k}, \\
&& \widehat{R}^{(q,0,\alpha)}_k = R^{(q,0,\alpha)}_a + R^{(q,0,\alpha)}_b + R^{(q,0,\alpha)}_c + R^{(q,0,\alpha)}_{g_k}.
\end{eqnarray*}
As in the case of the Minkowski background consider the integral of
(\ref{S:integrand_01}) over $\mathcal{N}_t$. Using the Gauss theorem on the
first two terms of equation (\ref{S:integrand_01}) one has that
\begin{eqnarray*}
&& \int_{\mathcal{N}_t} \partial_\tau \left( (1+\tau+a) |D^{q,0,\alpha}(\partial^p_\rho\phi_k)|^2 + (1-\tau-a) |D^{q,0,\alpha}(\partial^p_\rho\phi_{k+1})|^2 \right) \mbox{d} \tau \mbox{d} \rho \mbox{d} \mu \nonumber \\
&& \hspace{1cm} + \int_{\mathcal{N}_t} \partial_\rho  \left( (\rho+b) |D^{q,0,\alpha}(\partial^p_\rho\phi_{k+1})|^2 -(\rho+b) |D^{q,0,\alpha}(\partial^p_\rho\phi_k) |^2 \right) \mbox{d}\tau \mbox{d} \rho \mbox{d} \mu \nonumber \\
&& \hspace{2cm} = \int_{\mathcal{S}_t} \left( (1+t+a)|D^{q,0,\alpha}(\partial^p_\rho \phi_k)|^2 +(1-t-a) |D^{q,0,\alpha}(\partial^p_\rho \phi_{k+1})|^2  \right)\mbox{d} \rho \mbox{d} \mu \nonumber \\
&& \hspace{3cm}- \int_{\mathcal{S}_0} \left( |D^{q,0,\alpha}(\partial^p_\rho \phi_k)|^2 + |D^{q,0,\alpha}(\partial^p_\rho \phi_{k+1})|^2\right) \mbox{d}\rho \mbox{d}\mu \nonumber \\
&& \hspace{3cm}+ \int_{\mathcal{B}_t} \left( (1+\tau+a) n_\tau-(\rho+b) n_\rho)|D^{q,0,\alpha}\phi_k|^2 \right. \nonumber \\
&& \hspace{4cm} \left. +((1-\tau-a)n_\tau+(\rho+b) n_\rho)|D^{q,0,\alpha}\phi_{k+1}|^2 \right)\mbox{d}n \mbox{d} \mu \nonumber \\
&& \hspace{3cm}-\int_{\mathcal{I}_t} (\rho+b)\left(|D^{q,0,\alpha}\phi_{k+1}|^2-|D^{q,0,\alpha}\phi_k|^2 \right) \mbox{d}\tau \mbox{d} \mu,
\end{eqnarray*} 
with $n_\tau=\nu (\rho+b)$ and $n_\rho=\nu(1+\tau+a)$, and $\nu$ a normalisation factor. Note that in particular,
\[
\int_{\mathcal{I}_t} (\rho+b)\left(|D^{q,0,\alpha}\phi_{k+1}|^2-|D^{q,0,\alpha}\phi_k|^2 \right) \mbox{d}\tau \mbox{d} \mu=0,
\]
and that 
\[
\int_{\mathcal{B}_t} \left( (1+\tau+a) n_\tau-(\rho+b) n_\rho)|D^{q,0,\alpha}\phi_k|^2+((1-\tau-a)n_\tau+(\rho+b) n_\rho)|D^{q,0,\alpha}\phi_{k+1}|^2 \right)\mbox{d}n \mbox{d} \mu \geq 0.
\]
Hence, one has that
\begin{eqnarray}
&& \int_{\mathcal{S}_t} \bigg( (1+t+a)|D^{q,0,\alpha}(\partial^p_\rho\phi_k)|^2 + (1-t-a) |D^{q,0,\alpha}(\partial^p_\rho\phi_{k+1})|^2 \bigg) \mbox{d}\rho \mbox{d}\mu \nonumber \\
&& \hspace{1cm} +\int_{\mathcal{N}_t} (1+c)\left( D^{q,0,\alpha}(\partial^p_\rho \overline{\phi}_k) D^{q,0,\alpha} X_+(\partial^p_\rho\phi_{k+1}) + D^{q,0,\alpha} (\partial^p_\rho\phi_{k+1}) D^{q,0,\alpha}X_+(\partial^p_\rho\overline{\phi}_k) \right) \mbox{d}\tau \mbox{d}\rho \mbox{d}\mu \nonumber \\
&& \hspace{1cm} +\int_{\mathcal{N}_t} (1+c)\left(D^{q,0,\alpha}(\partial^p_\rho \phi_k D^{q,0,\alpha}) X_-(\partial^p_\rho\overline{\phi}_{k+1}) + D^{q,0,\alpha}(\partial^p_\rho\overline{\phi}_{k+1}) D^{q,0,\alpha} X_-(\partial^p_\rho \phi_k)\right) \mbox{d}\tau \mbox{d}\rho \mbox{d} \mu \nonumber \\
&& \hspace{1cm} +2\int_{\mathcal{N}_t}\bigg(p-q-k+g_k+\big(p-\frac{1}{2}\big)D^{0,1,0}b-\big(q-\frac{1}{2}\big)D^{1,0,0}a\bigg) |D^{q,0,\alpha}(\partial^p_\rho \phi_{k+1})|^2 \mbox{d}\tau \mbox{d} \rho \mbox{d}\mu \nonumber \\
&& \hspace{1cm} + \int_{\mathcal{N}_t} \left( D^{q,0,\alpha}(\partial^p_\rho \overline{\phi}_k) \widehat{F}^{(q,0,\alpha)}_k + D^{q,0,\alpha}(\partial^p_\rho \phi_k ) \overline{\widehat{F}}^{(q,0,\alpha)}_k\right) \mbox{d} \tau \mbox{d} \rho \mbox{d} \mu \nonumber \\
&& \hspace{1cm} + \int_{\mathcal{N}_t} \left( D^{q,0,\alpha}(\partial^p_\rho \overline{\phi}_k) \widehat{G}^{(q,0,\alpha)}_k + D^{q,0,\alpha}(\partial^p_\rho \phi_k ) \overline{\widehat{G}}^{(q,0,\alpha)}_k\right) \mbox{d} \tau \mbox{d} \rho \mbox{d} \mu \nonumber \\
&& \hspace{1cm} + \int_{\mathcal{N}_t} \left( D^{q,0,\alpha}(\partial^p_\rho \overline{\phi}_k) \widehat{S}^{(q,0,\alpha)}_k + D^{q,0,\alpha}(\partial^p_\rho \phi_k ) \overline{\widehat{S}}^{(q,0,\alpha)}_k\right) \mbox{d} \tau \mbox{d} \rho \mbox{d} \mu \nonumber \\
&& \hspace{1cm} + \int_{\mathcal{N}_t} \left( D^{q,0,\alpha}(\partial^p_\rho \overline{\phi}_k) \widehat{R}^{(q,0,\alpha)}_k + D^{q,0,\alpha}(\partial^p_\rho \phi_k ) \overline{\widehat{R}}^{(q,0,\alpha)}_k\right) \mbox{d} \tau \mbox{d} \rho \mbox{d} \mu \nonumber \\
&& \hspace{2cm} \leq \int_{\mathcal{S}_0} \big( |D^{q,0,\alpha}(\partial^p_\rho\phi_k|^2) + |D^{q,0,\alpha}(\partial^p_\rho\phi_{k+1})|^2\big) \mbox{d} \rho \mbox{d} \mu \nonumber \\
&& \hspace{3cm}+ 2 \int_{\mathcal{N}_t} \bigg(p-q+k-1-f_k+\big(p-\frac{1}{2}\big)D^{0,1,0}b-\big(q-\frac{1}{2}\big)D^{1,0,0}a\bigg)|D^{q,0,\alpha}(\partial^p_\rho \phi_k) |^2 \mbox{d} \tau \mbox{d} \rho \mbox{d} \mu. \nonumber \\
 \label{Schw_ineq_1}
\end{eqnarray}

In the sequel it shall be used that
\begin{eqnarray}
&& \hspace{-2.5cm}\left|\int_{\mathcal{N}_t} \left( D^{q,0,\alpha}(\partial^p_\rho \overline{\phi}_k) \widehat{F}_k + D^{q,0,\alpha}(\partial^p_\rho \phi_k ) \overline{\widehat{F}}_k\right) \mbox{d} \tau \mbox{d} \rho \mbox{d} \mu \right|\leq \int_{\mathcal{N}_t} \left( \varepsilon|D^{q,0,\alpha}(\partial^p_\rho \phi_k)|^2 + \frac{1}{\varepsilon}|\widehat{F}_k|^2 \right) \mbox{d}\tau \mbox{d} \rho \mbox{d} \mu, \label{S:auxiliary_03}\\
&& \hspace{-2.5cm} \left|\int_{\mathcal{N}_t} \left( D^{q,0,\alpha}(\partial^p_\rho \overline{\phi}_k) \widehat{G}_k + D^{q,0,\alpha}(\partial^p_\rho \phi_k ) \overline{\widehat{G}}_k\right) \mbox{d} \tau \mbox{d} \rho \mbox{d} \mu  \right| \leq \int_{\mathcal{N}_t} \left( \varepsilon|D^{q,0,\alpha}(\partial^p_\rho \phi_k)|^2 + \frac{1}{\varepsilon}|\widehat{G}_k|^2 \right) \mbox{d}\tau \mbox{d} \rho \mbox{d} \mu, \label{S:auxiliary_04}
\end{eqnarray}
for some small $\varepsilon>0$. Particular attention will be given to controlling the last two terms
in the left hand side of inequality (\ref{Schw_ineq_1}). To this end
choose $\rho_*$ such that
\[
|\widehat{S}^{(q,0,\alpha)}_k| \leq \delta, \quad |\widehat{R}^{(q,0,\alpha)}_k| \leq \delta,
\]
for $(\tau,\rho,\varsigma)\in \mathcal{N}_t$ and $\delta$ a suitably
small non-negative number: this can always be done as
$\widehat{S}^{(q,0,\alpha)}_k$ and $\widehat{R}^{(q,0,\alpha)}_k$ are
homogeneous functions of
$J^i(D^{j,p,\beta}\phi_k)$ and $J^i(D^{j,p,\beta}\phi_{k+1})$ with
$i\geq 1$ and $j+|\beta| \leq m$. Further, one has that 
\begin{eqnarray}
&& \hspace{-2cm} \left|\int_{\mathcal{N}_t} \left( D^{q,0,\alpha}(\partial^p_\rho \overline{\phi}_k) \widehat{S}^{(q,0,\alpha)}_k + D^{q,0,\alpha}(\partial^p_\rho \phi_k ) \overline{\widehat{S}}^{(q,0,\alpha)}_k\right) \mbox{d} \tau \mbox{d} \rho \mbox{d} \mu  \right| \nonumber \\
&& \hspace{3cm} \leq 2\int_{\mathcal{N}_t} \left|  D^{q,0,\alpha}(\partial^p_\rho \phi_k) \right| \left| \widehat{S}^{(q,0,\alpha)}_k  \right| \mbox{d} \tau \mbox{d} \rho \mbox{d} \mu \nonumber \\
&& \hspace{3cm} \leq2 \delta \int_{\mathcal{N}_t}  \left|  D^{q,0,\alpha}(\partial^p_\rho \phi_k) \right| \mbox{d} \tau \mbox{d} \rho \mbox{d} \mu \nonumber \\
&& \hspace{3cm} \leq 2\delta \int_{\mathcal{N}_t} \left( 1+|D^{q,0,\alpha}(\partial^p_\rho \phi_k)|^2\right) \mbox{d} \tau \mbox{d} \rho \mbox{d} \mu. \label{S:auxiliary_01}
\end{eqnarray}
Similarly, one finds that
\begin{equation}
\left|\int_{\mathcal{N}_t} \left( D^{q,0,\alpha}(\partial^p_\rho \overline{\phi}_k) \widehat{R}^{(q,0,\alpha)}_k + D^{q,0,\alpha}(\partial^p_\rho \phi_k ) \overline{\widehat{R}}^{(q,0,\alpha)}_k\right) \mbox{d} \tau \mbox{d} \rho \mbox{d} \mu  \right|  \leq 2\delta \int_{\mathcal{N}_t} \left( 1+|D^{q,0,\alpha}(\partial^p_\rho \phi_k)|^2\right) \mbox{d} \tau \mbox{d} \rho \mbox{d} \mu, \label{S:auxiliary_02}
\end{equation}
for suitable $\rho_*$. Using the inequalities (\ref{S:auxiliary_01}) and
(\ref{S:auxiliary_02}) together with (\ref{S:auxiliary_03}) and
(\ref{S:auxiliary_04}) one arrives at
\begin{eqnarray*}
&& \int_{\mathcal{S}_t} \bigg( (1+t+a)|D^{q,0,\alpha}(\partial^p_\rho\phi_k)|^2 + (1-t-a) |D^{q,0,\alpha}(\partial^p_\rho\phi_{k+1})|^2 \bigg) \mbox{d}\rho \mbox{d}\mu \nonumber \\
&& \hspace{0.5cm} +\int_{\mathcal{N}_t} (1+c)\left( D^{q,0,\alpha}(\partial^p_\rho \overline{\phi}_k) D^{q,0,\alpha} X_+(\partial^p_\rho\phi_{k+1}) + D^{q,0,\alpha} (\partial^p_\rho\phi_{k+1}) D^{q,0,\alpha}X_+(\partial^p_\rho\overline{\phi}_k) \right) \mbox{d}\tau \mbox{d}\rho \mbox{d}\mu \nonumber \\
&& \hspace{0.5cm} +\int_{\mathcal{N}_t} (1+c)\left(D^{q,0,\alpha}(\partial^p_\rho \phi_k) D^{q,0,\alpha} X_-(\partial^p_\rho\overline{\phi}_{k+1}) + D^{q,0,\alpha}(\partial^p_\rho\overline{\phi}_{k+1}) D^{q,0,\alpha} X_-(\partial^p_\rho \phi_k)\right) \mbox{d}\tau \mbox{d}\rho \mbox{d} \mu \nonumber \\
&& \hspace{0.5cm} +2\int_{\mathcal{N}_t}\bigg(p-q-k-\frac{\varepsilon}{2}-\delta +g_k+\big(p-\frac{1}{2}\big)D^{0,1,0}b-\big(q-\frac{1}{2}\big)D^{1,0,0}a \bigg) |D^{q,0,\alpha}(\partial^p_\rho \phi_{k+1})|^2 \mbox{d}\tau \mbox{d} \rho \mbox{d}\mu \nonumber \\
&& \hspace{1cm} \leq \int_{\mathcal{S}_0} \big( |D^{q,0,\alpha}(\partial^p_\rho\phi_k)|^2) + |D^{q,0,\alpha}(\partial^p_\rho\phi_{k+1})|^2\big) \mbox{d} \rho \mbox{d} \mu \nonumber \\
&& \hspace{1.5cm}+ 2 \int_{\mathcal{N}_t} \bigg(p-q+k-f_k-1+\frac{\varepsilon}{2}+\delta+\big(p-\frac{1}{2}\big)D^{0,1,0}b-\big(q-\frac{1}{2}\big)D^{1,0,0}a \bigg)|D^{q,0,\alpha}(\partial^p_\rho \phi_k) |^2 \mbox{d} \tau \mbox{d} \rho \mbox{d} \mu. \nonumber \\
&& \hspace{1.5cm} + \int_{\mathcal{N}_t} \left( 2\delta+ \frac{1}{\varepsilon}|\widehat{F}^{(q,0,\alpha)}_k|^2 + \frac{1}{\varepsilon}|\widehat{G}^{(q,0,\alpha)}_k|^2  \right) \mbox{d} \tau \mbox{d} \rho \mbox{d} \mu.
\end{eqnarray*}
At this point we note that
\begin{eqnarray*}
&& f_k=\mathcal{O}(\rho), \quad D^{1,0,0} a=\mathcal{O}(\rho), \\
&& g_k=\mathcal{O}(\rho), \quad D^{0,1,0}b =\mathcal{O}(\rho).
\end{eqnarray*}
Hence ---by making $\rho_*$ even smaller, if necessary--- one can make
\begin{eqnarray*}
&&\left| g_k +\left(p-\frac{1}{2} \right)D^{0,1,0}b -\left(q-\frac{1}{2}\right)D^{1,0,0}a \right|\leq \eta, \\
&&\left| -f_k +\left(p-\frac{1}{2} \right)D^{0,1,0}b -\left(q-\frac{1}{2}\right)D^{1,0,0}a \right|\leq \eta, 
\end{eqnarray*}
for suitably small $\eta>0$. Consequently one arrives to the inequality
\begin{eqnarray*}
&& \int_{\mathcal{S}_t} \bigg( (1+t+a)|D^{q,0,\alpha}(\partial^p_\rho\phi_k)|^2 + (1-t-a) |D^{q,0,\alpha}(\partial^p_\rho\phi_{k+1})|^2 \bigg) \mbox{d}\rho \mbox{d}\mu \nonumber \\
&& \hspace{1cm} +\int_{\mathcal{N}_t} (1+c)\left( D^{q,0,\alpha}(\partial^p_\rho \overline{\phi}_k) D^{q,0,\alpha} X_+(\partial^p_\rho\phi_{k+1}) + D^{q,0,\alpha} (\partial^p_\rho\phi_{k+1}) D^{q,0,\alpha}X_+(\partial^p_\rho\overline{\phi}_k) \right) \mbox{d}\tau \mbox{d}\rho \mbox{d}\mu \nonumber \\
&& \hspace{1cm} +\int_{\mathcal{N}_t}(1+c) \left(D^{q,0,\alpha}(\partial^p_\rho \phi_k) D^{q,0,\alpha} X_-(\partial^p_\rho\overline{\phi}_{k+1}) + D^{q,0,\alpha}(\partial^p_\rho\overline{\phi}_{k+1}) D^{q,0,\alpha} X_-(\partial^p_\rho \phi_k)\right) \mbox{d}\tau \mbox{d}\rho \mbox{d} \mu \nonumber \\
&& \hspace{1cm} +2\int_{\mathcal{N}_t}\bigg(p-q-k-\frac{\varepsilon}{2}-\delta -\eta\bigg) |D^{q,0,\alpha}(\partial^p_\rho \phi_{k+1})|^2 \mbox{d}\tau \mbox{d} \rho \mbox{d}\mu \nonumber \\
&& \hspace{2cm} \leq \int_{\mathcal{S}_0} \big( |D^{q,0,\alpha}(\partial^p_\rho\phi_k)|^2) + |D^{q,0,\alpha}(\partial^p_\rho\phi_{k+1})|^2\big) \mbox{d} \rho \mbox{d} \mu \nonumber \\
&& \hspace{3cm}+ 2 \int_{\mathcal{N}_t} \bigg(p-q+k-1+\frac{\varepsilon}{2}+\delta+\eta \bigg)|D^{q,0,\alpha}(\partial^p_\rho \phi_k) |^2 \mbox{d} \tau \mbox{d} \rho \mbox{d} \mu. \nonumber \\
&& \hspace{3cm} + \int_{\mathcal{N}_t} \left( 2\delta+ \frac{1}{\varepsilon}|\widehat{F}^{(q,0,\alpha)}_k|^2 + \frac{1}{\varepsilon}|\widehat{G}^{(q,0,\alpha)}_k|^2  \right) \mbox{d} \tau \mbox{d} \rho \mbox{d} \mu,
\end{eqnarray*} 
for suitably small $\rho_*$. The next step is to sum over $q$ and 
$\alpha$ for $q+|\alpha|\leq m$. One obtains
\begin{eqnarray*}
&& \sum_{q+|\alpha|\leq m} \int_{\mathcal{S}_t} \bigg( (1+t+a)|D^{q,0,\alpha}(\partial^p_\rho\phi_k)|^2 + (1-t-a) |D^{q,0,\alpha}(\partial^p_\rho\phi_{k+1})|^2 \bigg) \mbox{d}\rho \mbox{d}\mu \nonumber \\
&& \hspace{1cm} +2 \sum_{q+|\alpha|\leq m} \bigg(p-q-k-\frac{\varepsilon}{2}-\delta -\eta\bigg) \int_{\mathcal{N}_t} |D^{q,0,\alpha}(\partial^p_\rho \phi_{k+1})|^2 \mbox{d}\tau \mbox{d} \rho \mbox{d}\mu \nonumber \\
&& \hspace{2cm} \leq \sum_{q+|\alpha|\leq m} \int_{\mathcal{S}_0} \big( |D^{q,0,\alpha}(\partial^p_\rho\phi_k)|^2) + |D^{q,0,\alpha}(\partial^p_\rho\phi_{k+1})|^2\big) \mbox{d} \rho \mbox{d} \mu \nonumber \\
&& \hspace{3cm}+ 2\sum_{q+|\alpha|\leq m} \bigg(p-q+k-1+\frac{\varepsilon}{2}+\delta+\eta \bigg) \int_{\mathcal{N}_t} |D^{q,0,\alpha}(\partial^p_\rho \phi_k) |^2 \mbox{d} \tau \mbox{d} \rho \mbox{d} \mu. \nonumber \\
&& \hspace{3cm} + \sum_{q+|\alpha|\leq m} \int_{\mathcal{N}_t} \left( 2\delta+ \frac{1}{\varepsilon}|\widehat{F}^{(q,0,\alpha)}_k|^2 + \frac{1}{\varepsilon}|\widehat{G}^{(q,0,\alpha)}_k|^2  \right) \mbox{d} \tau \mbox{d} \rho \mbox{d} \mu,
\end{eqnarray*} 
where it has been used that
\begin{eqnarray*}
&& \sum_{q+|\alpha|\leq m} \int_{\mathcal{N}_t}(1+c) \left( D^{q,0,\alpha}(\partial^p_\rho \overline{\phi}_k) D^{q,0,\alpha} X_+(\partial^p_\rho\phi_{k+1}) + D^{q,0,\alpha} (\partial^p_\rho\phi_{k+1}) D^{q,0,\alpha}X_+(\partial^p_\rho\overline{\phi}_k) \right) \mbox{d}\tau \mbox{d}\rho \mbox{d}\mu =0 \\
&& \sum_{q+|\alpha|\leq m} \int_{\mathcal{N}_t}(1+c) \left(D^{q,0,\alpha}(\partial^p_\rho \phi_k D^{q,0,\alpha}) X_-(\partial^p_\rho\overline{\phi}_{k+1}) + D^{q,0,\alpha}(\partial^p_\rho\overline{\phi}_{k+1}) D^{q,0,\alpha} X_-(\partial^p_\rho \phi_k)\right) \mbox{d}\tau \mbox{d}\rho \mbox{d} \mu =0,
\end{eqnarray*}
as a consequence of lemma \ref{lemma:Liegroups} in appendix
\ref{appendix:SU(2)} and of $Xc=0$, $X_\pm c=0$. Note that for $k=0,1$
one has that
\begin{eqnarray*}
&& 2 \sum_{q+|\alpha|\leq m} \left(p-q-k-\frac{\varepsilon}{2}-\delta -\eta\right) \int_{\mathcal{N}_t} |D^{q,0,\alpha}(\partial^p_\rho \phi_{k+1})|^2 \mbox{d}\tau \mbox{d} \rho \mbox{d}\mu  \nonumber \\
&& \hspace{3cm} \geq  2 \left(p-m-1-\frac{\varepsilon}{2}-\delta -\eta \right) \sum_{q+|\alpha|\leq m} \int_{\mathcal{N}_t} |D^{q,0,\alpha}(\partial^p_\rho \phi_{k+1})|^2 \mbox{d}\tau \mbox{d} \rho \mbox{d}\mu, \nonumber \\
&&  2\sum_{q+|\alpha|\leq m} \left(p-q+k-1+\frac{\varepsilon}{2}+\delta+\eta \right) \int_{\mathcal{N}_t} |D^{q,0,\alpha}(\partial^p_\rho \phi_k) |^2 \mbox{d} \tau \mbox{d} \rho \mbox{d} \mu \nonumber \\
&& \hspace{3cm} \leq 2\left(p+ \frac{\varepsilon}{2}+\delta+\eta \right) \sum_{q+|\alpha|\leq m} \int_{\mathcal{N}_t} |D^{q,0,\alpha}(\partial^p_\rho \phi_k) |^2 \mbox{d} \tau \mbox{d} \rho \mbox{d} \mu,
\end{eqnarray*}
from where it follows that
\begin{eqnarray}
&& \sum_{q+|\alpha|\leq m} \int_{\mathcal{S}_t} \bigg( (1+t+a)|D^{q,0,\alpha}(\partial^p_\rho\phi_k)|^2 + (1-t-a) |D^{q,0,\alpha}\partial^p_\rho(\phi_{k+1})|^2 \bigg) \mbox{d}\rho \mbox{d}\mu \nonumber \\
&& \hspace{1cm} +2  \left(p-m-1-\frac{\varepsilon}{2}-\delta -\eta \right) \sum_{q+|\alpha|\leq m}  \int_{\mathcal{N}_t} |D^{q,0,\alpha}(\partial^p_\rho \phi_{k+1})|^2 \mbox{d}\tau \mbox{d} \rho \mbox{d}\mu \nonumber \\
&& \hspace{2cm} \leq \sum_{q+|\alpha|\leq m} \int_{\mathcal{S}_0} \big( |D^{q,0,\alpha}(\partial^p_\rho\phi_k|^2) + |D^{q,0,\alpha}(\partial^p_\rho\phi_{k+1})|^2\big) \mbox{d} \rho \mbox{d} \mu \nonumber \\
&& \hspace{3cm}+ 2 \left(p+ \frac{\varepsilon}{2}+\delta+\eta \right) \sum_{q+|\alpha|\leq m}  \int_{\mathcal{N}_t} |D^{q,0,\alpha}(\partial^p_\rho \phi_k) |^2 \mbox{d} \tau \mbox{d} \rho \mbox{d} \mu \nonumber \\
&& \hspace{3cm} + \sum_{q+|\alpha|\leq m} \int_{\mathcal{N}_t} \left( 2\delta+ \frac{1}{\varepsilon}|\widehat{F}^{(q,0,\alpha)}_k|^2 + \frac{1}{\varepsilon}|\widehat{G}^{(q,0,\alpha)}_k|^2  \right) \mbox{d} \tau \mbox{d} \rho \mbox{d} \mu. \label{S:inequality_03}
\end{eqnarray} 
In order to guarantee that the second term on the left hand side of
inequality (\ref{S:inequality_03}) is positive ---so that it can be
removed from the inequality---, choose $p$ such that $p > m+1 +\varepsilon/2 + \delta
+ \eta$. Hence $\rho_*$ will be suitably chosen such that
$\varepsilon/2 + \delta + \eta <1$. Further, for $t\in[0,1]$ it holds that
\[
\int_{\mathcal{S}_t}|D^{q,0,\alpha}(\partial^p_\rho\phi_k)|^2 \mbox{d}\rho \mbox{d}\mu \leq \int_{\mathcal{S}_t} \bigg( (1+t+a)|D^{q,0,\alpha}(\partial^p_\rho\phi_k)|^2 + (1-t-a) |D^{q,0,\alpha}\partial^p_\rho(\phi_{k+1})|^2 \bigg) \mbox{d}\rho \mbox{d}\mu,
\]
for suitably small $\rho_*>0$ such that $(1+t+a)\geq 1$ on
$\mathcal{S}_t$ ---note that as discussed in section \ref{F_gauge},
equation (\ref{c0_scri_1}) $(\tau+a)\big|_{\scri^+} =1$--- and
$(1-t-a)\geq 0$. Hence one arrives to the inequality
\begin{eqnarray}
&& \sum_{q+|\alpha|\leq m} \int_{\mathcal{S}_t} |D^{q,0,\alpha}(\partial^p_\rho\phi_k)|^2  \mbox{d}\rho \mbox{d}\mu \leq \sum_{q+|\alpha|\leq m} \int_{\mathcal{S}_0} \big( |D^{q,0,\alpha}(\partial^p_\rho\phi_k|^2) + |D^{q,0,\alpha}(\partial^p_\rho\phi_{k+1})|^2\big) \mbox{d} \rho \mbox{d} \mu \nonumber \\
&& \hspace{6cm}+ 2 (p+ 1) \sum_{q+|\alpha|\leq m}  \int_{\mathcal{N}_t} |D^{q,0,\alpha}(\partial^p_\rho \phi_k) |^2 \mbox{d} \tau \mbox{d} \rho \mbox{d} \mu \nonumber \\
&& \hspace{6cm} + \sum_{q+|\alpha|\leq m} \int_{\mathcal{N}_t} \left( 2\delta+ \frac{1}{\varepsilon}|\widehat{F}^{(q,0,\alpha)}_k|^2 + \frac{1}{\varepsilon}|\widehat{G}^{(q,0,\alpha)}_k|^2  \right) \mbox{d} \tau \mbox{d} \rho \mbox{d} \mu, \nonumber \\
&& \label{schw_ineq_2}
\end{eqnarray} 
for $p>m+2$ for suitably small $\rho_*$ and $k=0,1$. Note also that 
\begin{eqnarray}
&& 2 (p-m-2) \sum_{q+|\alpha|\leq m}  \int_{\mathcal{N}_t} |D^{q,0,\alpha}(\partial^p_\rho \phi_{k+1})|^2 \mbox{d}\tau \mbox{d} \rho \mbox{d}\mu \nonumber \\
&& \hspace{2cm} \leq \sum_{q+|\alpha|\leq m} \int_{\mathcal{S}_0} \big( |D^{q,0,\alpha}(\partial^p_\rho\phi_k)|^2) + |D^{q,0,\alpha}(\partial^p_\rho\phi_{k+1})|^2\big) \mbox{d} \rho \mbox{d} \mu \nonumber \\
&& \hspace{3cm}+ 2(p+1) \sum_{q+|\alpha|\leq m}  \int_{\mathcal{N}_t} |D^{q,0,\alpha}(\partial^p_\rho \phi_k) |^2 \mbox{d} \tau \mbox{d} \rho \mbox{d} \mu \nonumber \\
&& \hspace{3cm} + \sum_{q+|\alpha|\leq m} \int_{\mathcal{N}_t} \left( 2\delta+ \frac{1}{\varepsilon}|\widehat{F}^{(q,0,\alpha)}_k|^2 + \frac{1}{\varepsilon}|\widehat{G}^{(q,0,\alpha)}_k|^2  \right) \mbox{d} \tau \mbox{d} \rho \mbox{d} \mu, \label{schw_ineq_3}
\end{eqnarray}  
for $t\in[0,1]$ and suitably small $\rho_*$. Applying the standard
Gronwall argument to inequality (\ref{schw_ineq_2}) one gets
\begin{eqnarray}
&&  \sum_{q+|\alpha|\leq m} \int_{\mathcal{N}_t} |D^{q,0,\alpha}(\partial^p_\rho\phi_k)|^2  \mbox{d}\tau \mbox{d}\rho \mbox{d}\mu \nonumber \\
&& \hspace{2cm} \leq\frac{1}{2(p+1)}\left( e^{2(p+1)t}-1 \right) \sum_{q+|\alpha|\leq m} \int_{\mathcal{S}_0} \big( |D^{q,0,\alpha}(\partial^p_\rho\phi_k)|^2) + |D^{q,0,\alpha}(\partial^p_\rho\phi_{k+1})|^2\big) \mbox{d} \rho \mbox{d} \mu \nonumber \\
&&  \hspace{3cm} +e^{2(p+1)t} \int^t_0 e^{-2(p+1)s} \left( \sum_{q+|\alpha|\leq m} \int_{\mathcal{N}_t} \left( 2\delta+ \frac{1}{\varepsilon}|\widehat{F}^{(q,0,\alpha)}_k|^2 + \frac{1}{\varepsilon}|\widehat{G}^{(q,0,\alpha)}_k|^2  \right) \mbox{d} \tau \mbox{d} \rho \mbox{d} \mu \right)   \mbox{d} s, \nonumber \\
&& \label{schw_ineq_4}
\end{eqnarray}
for $k=0,1$, $t\in[0,1]$, $p>m+2$ and suitably small $\rho_*$. Note
that the second term in the right hand side of the last inequality is
bounded for $t\in[0,1]$. In order to obtain an estimate for
$D^{q,0,\alpha}(\partial^p_\rho \phi_2)$, use inequality
(\ref{schw_ineq_3}) in conjunction with (\ref{schw_ineq_4}) so that
one obtains:
\begin{eqnarray*}
&& \sum_{q+|\alpha|\leq m} \int_{\mathcal{N}_t} |D^{q,0,\alpha}(\partial^p_\rho\phi_2)|^2  \mbox{d}\tau \mbox{d}\rho \mbox{d}\mu \nonumber \\
&& \hspace{1cm}  \leq \frac{e^{2(p+1)t}}{2(p-m-2)}   \sum_{q+|\alpha|\leq m}  \int_{\mathcal{S}_0} \big( |D^{q,0,\alpha}(\partial^p_\rho\phi_1)|^2) + |D^{q,0,\alpha}(\partial^p_\rho\phi_{2})|^2\big) \mbox{d} \rho \mbox{d} \mu \nonumber \\
&& \hspace{2cm}  + \frac{2(p+1)}{2(p-m-2)} e^{2(p+1)t} \int^t_0 e^{-2(p+1)s} \left( \sum_{q+|\alpha|\leq m} \int_{\mathcal{N}_t} \left( 2\delta+ \frac{1}{\varepsilon}|\widehat{F}^{(q,0,\alpha)}_1|^2 + \frac{1}{\varepsilon}|\widehat{G}^{(q,0,\alpha)}_1|^2  \right) \mbox{d} \tau \mbox{d} \rho \mbox{d} \mu \right)   \mbox{d} s \\
&& \hspace{2cm} + \frac{1}{2(p-m-2)}  \sum_{q+|\alpha|\leq m} \int_{\mathcal{N}_t} \left( 2\delta+ \frac{1}{\varepsilon}|\widehat{F}^{(q,0,\alpha)}_1|^2 + \frac{1}{\varepsilon}|\widehat{G}^{(q,0,\alpha)}_1|^2  \right) \mbox{d} \tau \mbox{d} \rho \mbox{d} \mu,
\end{eqnarray*}
for $p>m+2$. Again the last two terms on the right hand side of this last
inequality are bounded at $\tau=1$. Hence, given $p>m+2$ one can conclude that
there is a $\rho_*>0$, and positive constants $C_1(\rho_*)$ and $C_2(\rho_*)$
depending on $\rho_*$ such that
\begin{eqnarray}
&& \int_{\mathcal{N}_t} \left( \sum_{q+|\alpha|\leq m}  |D^{q,0,\alpha}(\partial^p_\rho \phi_k)|^2  \right) \mbox{d} \tau \mbox{d} \rho \mbox{d} \mu \nonumber \\
&& \hspace{2cm}\leq  C_1(\rho_*) \sum_{k=0}^2 \int_{\mathcal{S}_0} \left(
  \sum_{q+|\alpha|\leq m}  |D^{q,0,\alpha}(\partial^p_\rho \phi_k)|^2
\right)  \mbox{d} \rho \mbox{d} \mu +C_2(\rho_*). \label{base_estimates}
\end{eqnarray}
This inequality will be used as the base step in a finite inductive argument.

\subsection{Estimates for $D^{q,j,\alpha}(\partial^p_\rho \phi_k)$ and 
  $D^{q,j,\alpha}(\partial^p_\rho \phi_{k+1})$, $q+j+|\alpha|\leq m$} \label{section:general_step}

One can construct estimates for $D^{q,p',\alpha}(\partial^p_\rho
\phi_k)$, $D^{q,p',\alpha}(\partial^p_\rho \phi_{k+1})$,
$q+p'+|\alpha|\leq m$, where $m$ is an arbitrary, but fixed
non-negative integer by means of an inductive argument on $p'$. The
base step ($p'=0$) of this induction argument is given by the estimate
(\ref{base_estimates}) obtained in the previous subsection. 
Accordingly, it shall be assumed that there exist a suitably small
$\rho_*>0$ and constants $C_1(\rho_*)$ and $C_2(\rho_*)$ depending on
$\rho_*$ for which the quantities $D^{q,j,\alpha}(\partial^p_\rho
\phi_k)$, $D^{q,j,\alpha}(\partial^p_\rho \phi_{k+1})$, with $0\leq
j\leq p'-1<m$ satisfy estimates of the form
\begin{eqnarray}
&&  \int_{\mathcal{N}_t} \sum_{q+j+|\alpha|\leq m \atop 0\leq j <p'}
|D^{q,j,\alpha}(\partial^p_\rho \phi_k)|^2 \mbox{d}\tau \mbox{d}\rho \mbox{d}
\mu \nonumber \\
&&\hspace{2cm} \leq C_1(\rho_*) 
\sum^2_{k=0} \int_{\mathcal{S}_0} \left( \sum_{q+j+|\alpha|\leq m \atop 0 \leq
    j <p'}
  |D^{q,j,\alpha}(\partial^p_\rho \phi_k)|^2 \right) 
\mbox{d}\rho \mbox{d} \mu + C_2(\rho_*),  \label{induction_hypothesis_estimate}
\end{eqnarray}
for $k=0,1,2$. It will be shown that analogous estimates hold for
$0\leq j \leq p'\leq m$.

\bigskip
Using the expressions (\ref{DA}) and (\ref{DB}) one can rewrite
\[
D^{q,p+j,\alpha}\overline{\phi}_k D^{q,p+j,\alpha} A_k + D^{q,p+j,\alpha} \phi_k D^{q,p+j,\alpha}\overline{A}_k+ D^{q,p+j,\alpha}\overline{\phi}_k D^{q,p+j,\alpha} B_k + D^{q,p+j,\alpha} \phi_k D^{q,p+j,\alpha}\overline{B}_k=0, 
\]
for $k=0,1$ and $0\leq j\leq m$ as
\begin{eqnarray*}
&& 0= \partial_\tau \left( (1+\tau+a) |D^{q,j,\alpha}(\partial^p_\rho\phi_k)|^2 + (1-\tau-a) |D^{q,j,\alpha}(\partial^p_\rho\phi_{k+1})|^2 \right) \nonumber \\
&& \hspace{1cm}+ \partial_\rho  \left( (\rho+b) |D^{q,j,\alpha}(\partial^p_\rho\phi_{k+1})|^2 -(\rho+b) |D^{q,j,\alpha}(\partial^p_\rho\phi_k) |^2 \right) \nonumber \\
&& \hspace{1cm} +(1+c)\left( D^{q,j,\alpha} (\partial^p_\rho\overline{\phi}_k) D^{q,j,\alpha} X_+(\partial^p_\rho\phi_{k+1}) + D^{q,j,\alpha}(\partial^p_\rho\phi_{k+1}) D^{q,j,\alpha}X_+(\partial^p_\rho\overline{\phi}_k) \right)  \nonumber \\
&& \hspace{1cm} + (1+c)\left(D^{q,j,\alpha}(\partial^p_\rho\phi_k) D^{q,j,\alpha}X_-(\partial^p_\rho\overline{\phi}_{k+1}) + D^{q,j,\alpha}(\partial^p_\rho\overline{\phi}_{k+1}) D^{q,j,\alpha} X_- (\partial^p_\rho\phi_k)\right) \nonumber \\
&& \hspace{1cm} -2\left(p+j-q+k-1-f_k +\left(p-\frac{1}{2}\right)D^{0,1,0}b -\left(q-\frac{1}{2}\right)D^{1,0,0}a\right)|D^{q,j,\alpha}(\partial^p_\rho \phi_k)|^2 \nonumber \\
&& \hspace{1cm}+ 2\left(p+j-q-k+g_k +\left(p-\frac{1}{2}\right) D^{0,1,0}b -\left(q-\frac{1}{2}\right)D^{1,0,0}a\right) |D^{q,j,\alpha}(\partial^p_\rho \phi_{k+1})|^2. \nonumber \\
&& \hspace{1cm}+ D^{q,j,\alpha}(\partial^p_\rho\overline{\phi}_k) H^{(q,j,\alpha)}_a + D^{q,j,\alpha}(\partial^p_\rho\phi_k) \overline{H}^{(q,j,\alpha)}_a+ D^{q,j,\alpha}(\partial^p_\rho\overline{\phi}_k) H^{(q,j,\alpha)}_b + D^{q,j,\alpha}(\partial^p_\rho\phi_k) \overline{H}^{(q,j,\alpha)}_b \nonumber \\
&& \hspace{1cm}+ D^{q,j,\alpha}(\partial^p_\rho\overline{\phi}_k) H^{(q,j,\alpha)}_c + D^{q,j,\alpha}(\partial^p_\rho\phi_k) \overline{H}^{(q,j,\alpha)}_c+ D^{q,j,\alpha}(\partial^p_\rho\overline{\phi}_k) H^{(q,j,\alpha)}_{f_k} + D^{q,j,\alpha}(\partial^p_\rho\phi_k) \overline{H}^{(q,j,\alpha)}_{f_k} \nonumber\\
&& \hspace{1cm}+ D^{q,j,\alpha}(\partial^p_\rho\overline{\phi}_k) K^{(q,j,\alpha)}_a + D^{q,j,\alpha}(\partial^p_\rho\phi_k) \overline{K}^{(q,j,\alpha)}_a+ D^{q,j,\alpha}(\partial^p_\rho\overline{\phi}_k) K^{(q,j,\alpha)}_b + D^{q,j,\alpha}(\partial^p_\rho\phi_k) \overline{K}^{(q,j,\alpha)}_b \nonumber \\
&& \hspace{1cm}+ D^{q,j,\alpha}(\partial^p_\rho\overline{\phi}_k) K^{(q,j,\alpha)}_c + D^{q,j,\alpha}(\partial^p_\rho\phi_k) \overline{K}^{(q,j,\alpha)}_c+ D^{q,j,\alpha}(\partial^p_\rho\overline{\phi}_k) K^{(q,j,\alpha)}_{g_k} + D^{q,j,\alpha}(\partial^p_\rho\phi_k) \overline{K}^{(q,j,\alpha)}_{g_k}, \nonumber \\
&&
\end{eqnarray*}
with $H^{(q,j,\alpha)}_a$, $H^{(q,j,\alpha)}_b$, $H^{(q,j,\alpha)}_c$,
$H^{(q,j,\alpha)}_{f_k}$ and $K^{(q,j,\alpha)}_a$, $K^{(q,j,\alpha)}_b$,
$K^{(q,j,\alpha)}_c$, $K^{(q,j,\alpha)}_{g_k}$ given by formulae
(\ref{H_a})-(\ref{K_g}). As in subsection \ref{section:base_step}, a detailed
analysis of these terms will be crucial. For example, one has that
\begin{eqnarray*}
&& H^{(q,j,\alpha)}_a = \sum_{s=2}^q \binom{q}{s}D^{s,0,0}a D^{q-s+1,0,\alpha}\phi_k + \sum^q_{s=1}\sum^{p+j}_{l=1} \binom{q}{s} \binom{p+j}{l}D^{s,l,0}a D^{q-s,p+j-l,\alpha}\phi_k \\
&& \phantom{H^{(q,j,\alpha)}}= \sum_{s=2}^q \binom{q}{s}D^{s,0,0}a D^{q-s+1,0,\alpha}\phi_k   \\
&& \hspace{2.5cm} + \sum_{s=1}^q \sum_{l=1}^{j} \binom{q}{s} \binom{p+j}{l} D^{s,l,0}a D^{q-s,p+j-l,\alpha} \phi_k \\
&&  \hspace{2.5cm} + \sum_{s=1}^q \sum_{l=j+1}^{p+j} \binom{q}{s} \binom{p+j}{l} D^{s,l,0}a D^{q-s,p+j-l,\alpha} \phi_k.
\end{eqnarray*}
The terms in the second and fourth lines of the last formula contain at most $\rho$-derivatives of $\phi_k$ of order $p-1$. Thus, they can be handled as
in subsection \ref{section:base_step} by using the expansion Ansatz
(\ref{Ansatz}). Note however, that this approach is not applicable to the
expression in the third line as it contains $\rho$-derivatives of $\phi_k$ of
order $p$ and higher. These terms will by controlled by means of the induction
hypothesis for $0\leq j \leq p'-1 <m$. Using the expansion Ansatz
(\ref{Ansatz}) one has then that
\begin{eqnarray*}
&& H^{(q,j,\alpha)}_a = \sum^q_{s=2} \sum^{p-1}_{l=0} \frac{1}{l!} \binom{q}{s}
D^{s,0,0}aD^{q-s+1,0,\alpha} \phi^{(l)}_k \rho^l  \\
&& \hspace{2.5cm} + \sum^q_{s=2} \binom{q}{s} D^{s,0,0} a J^p(D^{q-s+1,0,\alpha} \phi_k) \\
&& \hspace{2.5cm} + U^{(q,j,\alpha)}_a \\
&& \hspace{2.5cm} + \sum^q_{s=1} \sum ^{p+j}_{l=j+1} \sum^{p-1}_{t=p+j-l} \frac{1}{(t-p-j+l)!}\binom{q}{s} \binom{p+j}{l}D^{s,l,\alpha}a D^{q-s,0,\alpha}\phi^{(t)}_k \rho ^{t-p-j+l} \\
&& \hspace{2.5cm} + \sum^q_{s=1} \sum_{l=j+1}^{p+j} \binom{q}{s} \binom{p+j}{l} D^{s,l,0} a J^{l-j} (D^{q-s,p,\alpha} \phi_k),
\end{eqnarray*}
for $0\leq j \leq p'\leq m$ with
\begin{equation}
U_a^{(q,j,\alpha)}= \sum^q_{s=1} \sum^{j}_{l=1} \binom{q}{s} \binom{p+j}{l} D^{s,l,0}a D^{q-s,p+j-l,\alpha} \phi_k. \label{definition:U}
\end{equation}
Hence, $H_a^{(q,j,\alpha)}$, $0\leq j \leq p'-1 <m$,  can be split as
\[
H_a^{(q,p,\alpha)} = F_a^{(q,j,\alpha)} + S_a^{(q,j,\alpha)}+ U^{(q,j,\alpha)}_a,
\]
with
\begin{eqnarray*}
&& F_a^{(q,j,\alpha)}=\sum^q_{s=2} \sum^{p-1}_{l=0} \frac{1}{l!} \binom{q}{s}
D^{s,0,0}aD^{q-s+1,0,\alpha} \phi^{(l)}_k \rho^l \\
&& \hspace{2.5cm}  + \sum^q_{s=1} \sum ^{p+j}_{l=j+1} \sum^{p-1}_{t=p+j-l} \frac{1}{(t-p-j+l)!}\binom{q}{s} \binom{p+j}{l}D^{s,l,\alpha}a D^{q-s,0,\alpha}\phi^{(t)}_k \rho ^{t-p-j+l},
\end{eqnarray*}
and
\begin{eqnarray*}
&& S_a^{(q,j,\alpha)} =  \sum^q_{s=2} \binom{q}{s} D^{s,0,0} a J^p(D^{q-s+1,0,\alpha} \phi_k)  \nonumber \\
&& \hspace{2.5cm}  + \sum^q_{s=1} \sum_{l=j+1}^{p+j} \binom{q}{s} \binom{p+j}{l} D^{s,l,0} a J^{l-j} (D^{q-s,p,\alpha} \phi_k).
\end{eqnarray*}
As in the discussion of section \ref{section:base_step}, the terms
$F_a^{(q,j,\alpha)}$, $0\leq j \leq p'\leq m$ can be calculated
explicitly using the transport equations on $\mathcal{I}$. The terms $S_a^{(q,j,\alpha)}$ are homogeneous in
$J^l(D^{i,n,\beta}\phi_k)$ with $l\geq 1$, $i+n+|\beta|\leq m$; hence
they can be controlled by choosing a suitable $\rho_*$. The term
$U^{(q,j,\alpha})$ contains $\rho$-derivatives of $\phi_k$ up to order
$p+j-1$, accordingly it will be controlled using the induction
hypothesis. A similar split can be introduced for the other $H$'s and
$K$'s. Hence, one is led to analyse the expression:
\begin{eqnarray}
&& 0= \partial_\tau \left( (1+\tau+a) |D^{q,j,\alpha}(\partial^p_\rho\phi_k)|^2 + (1-\tau-a) |D^{q,j,\alpha}(\partial^p_\rho\phi_{k+1})|^2 \right) \nonumber \\
&& \hspace{1cm}+ \partial_\rho  \left( (\rho+b) |D^{q,j,\alpha}(\partial^p_\rho\phi_{k+1})|^2 -(\rho+b) |D^{q,j,\alpha}(\partial^p_\rho\phi_k) |^2 \right) \nonumber \\
&& \hspace{1cm} +(1+c)\left( D^{q,j,\alpha} (\partial^p_\rho\overline{\phi}_k) D^{q,j,\alpha} X_+(\partial^p_\rho\phi_{k+1}) + D^{q,j,\alpha}(\partial^p_\rho\phi_{k+1}) D^{q,j,\alpha}X_+(\partial^p_\rho\overline{\phi}_k) \right)  \nonumber \\
&& \hspace{1cm} + (1+c)\left(D^{q,j,\alpha}(\partial^p_\rho\phi_k) D^{q,j,\alpha}X_-(\partial^p_\rho\overline{\phi}_{k+1}) + D^{q,j,\alpha}(\partial^p_\rho\overline{\phi}_{k+1}) D^{q,j,\alpha} X_- (\partial^p_\rho\phi_k)\right) \nonumber \\
&& \hspace{1cm} -2\left(p+j-q+k-1-f_k +\left(p-\frac{1}{2}\right)D^{0,1,0}b -\left(q-\frac{1}{2}\right)D^{1,0,0}a\right)|D^{q,j,\alpha}(\partial^p_\rho \phi_k)|^2 \nonumber \\
&& \hspace{1cm}+ 2\left(p+j-q-k+g_k +\left(p-\frac{1}{2}\right) D^{0,1,0}b -\left(q-\frac{1}{2}\right)D^{1,0,0}a\right) |D^{q,j,\alpha}(\partial^p_\rho \phi_{k+1})|^2. \nonumber \\
&& \hspace{1cm}+ D^{q,j,\alpha}(\partial^p_\rho\overline{\phi}_k) \widehat{F}^{(q,j,\alpha)}_k + D^{q,j,\alpha}(\partial^p_\rho\phi_k) \overline{\widehat{F}}^{(q,j,\alpha)}_k+ D^{q,j,\alpha}(\partial^p_\rho\overline{\phi}_k) \widehat{G}^{(q,j,\alpha)}_k + D^{q,j,\alpha}(\partial^p_\rho\phi_k) \overline{\widehat{G}}^{(q,j,\alpha)}_k \nonumber \\
&& \hspace{1cm}+ D^{q,j,\alpha}(\partial^p_\rho\overline{\phi}_k) \widehat{S}^{(q,j,\alpha)}_k + D^{q,j,\alpha}(\partial^p_\rho\phi_k) \overline{\widehat{S}}^{(q,j,\alpha)}_k+ D^{q,j,\alpha}(\partial^p_\rho\overline{\phi}_k) \widehat{R}^{(q,j,\alpha)}_k + D^{q,j,\alpha}(\partial^p_\rho\phi_k) \overline{\widehat{R}}^{(q,j,\alpha)}_k \nonumber\\
&& \hspace{1cm}+ D^{q,j,\alpha}(\partial^p_\rho\overline{\phi}_k) \widehat{U}^{(q,j,\alpha)}_k + D^{q,j,\alpha}(\partial^p_\rho\phi_k) \overline{\widehat{U}}^{(q,j,\alpha)}_k+ D^{q,j,\alpha}(\partial^p_\rho\overline{\phi}_k) \widehat{V}^{(q,j,\alpha)}_k + D^{q,j,\alpha}(\partial^p_\rho\phi_k) \overline{\widehat{V}}^{(q,j,\alpha)}_k, \nonumber \\
&& \label{general:integrand} 
\end{eqnarray} 
with
\begin{eqnarray*}
&& \widehat{F}^{(q,j,\alpha)}_k = F_a^{(q,j,\alpha)}+F_b^{(q,j,\alpha)}+F_c^{(q,j,\alpha)}+F_{f_k}^{(q,j,\alpha)}, \\
&& \widehat{G}_k^{(q,j,\alpha)} = G_a^{(q,j,\alpha)}+G_b^{(q,j,\alpha)}+G_c^{(q,j,\alpha)}+G_{g_k}^{(q,j,\alpha)}, \\
&& \widehat{R}_k^{(q,j,\alpha)} = R_a^{(q,j,\alpha)}+R_b^{(q,j,\alpha)}+R_c^{(q,j,\alpha)}+R_{f_k}^{(q,j,\alpha)}, \\
&& \widehat{S}_k^{(q,j,\alpha)} = S_a^{(q,j,\alpha)}+S_b^{(q,j,\alpha)}+S_c^{(q,j,\alpha)}+S_{g_k}^{(q,j,\alpha)}, \\
&& \widehat{U}_k^{(q,j,\alpha)} = U_a^{(q,j,\alpha)}+U_b^{(q,j,\alpha)}+U_c^{(q,j,\alpha)}+U_{f_k}^{(q,j,\alpha)}, \\
&& \widehat{V}_k^{(q,j,\alpha)} = V_a^{(q,j,\alpha)}+V_b^{(q,j,\alpha)}+V_c^{(q,j,\alpha)}+V_{g_k}^{(q,j,\alpha)},
\end{eqnarray*}
for $k=0,1$ and $q+j+|\alpha|\leq m$. The terms $V_a^{(q,j,\alpha)}$, $V_b^{(q,j,\alpha)}$, $V_c^{(q,j,\alpha)}$, $V_{g_k}^{(q,j,\alpha)}$ are defined in analogy to the expression for $U_a^{(q,j,\alpha)}$ given by formulae (\ref{definition:U}) and contain $\rho$-derivatives of $\phi_{k+1}$ up to order $p+j-1$.  

Following the ideas of the argument given in sections
\ref{section:Minkowski} and \ref{section:base_step}, integrating
(\ref{general:integrand}) over $\mathcal{N}_t$ and then using the Gauss
theorem on the first two terms one obtains the inequality
\begin{eqnarray*}
&&  \int_{\mathcal{S}_t} \left( (1+t+a)|D^{q,j,\alpha}(\partial^p_\rho \phi_k)|^2 +(1-t-a) |D^{q,j,\alpha}(\partial^p_\rho \phi_{k+1})|^2  \right)\mbox{d} \rho \mbox{d} \mu \nonumber \\
&& \hspace{0.5cm} +\int_{\mathcal{N}_t}(1+c)\left( D^{q,j,\alpha} (\partial^p_\rho\overline{\phi}_k) D^{q,j,\alpha} X_+(\partial^p_\rho\phi_{k+1}) + D^{q,j,\alpha}(\partial^p_\rho\phi_{k+1}) D^{q,j,\alpha}X_+(\partial^p_\rho\overline{\phi}_k) \right) \mbox{d} \tau \mbox{d}\rho \mbox{d} \mu \nonumber \\
&& \hspace{0.5cm} + \int_{\mathcal{N}_t} (1+c)\left(D^{q,j,\alpha}(\partial^p_\rho\phi_k) D^{q,j,\alpha}X_-(\partial^p_\rho\overline{\phi}_{k+1}) + D^{q,j,\alpha}(\partial^p_\rho\overline{\phi}_{k+1}) D^{q,j,\alpha} X_- (\partial^p_\rho\phi_k)\right) \mbox{d} \tau \mbox{d} \rho \mbox{d} \mu \nonumber \\
&& \hspace{0.5cm}+ 2\int_{\mathcal{N}_t}\left(p+j-q-k+g_k +\left(p-\frac{1}{2}\right) D^{0,1,0}b -\left(q-\frac{1}{2}\right)D^{1,0,0}a\right) |D^{q,j,\alpha}(\partial^p_\rho \phi_{k+1})|^2 \mbox{d}\tau \mbox{d} \rho \mbox{d} \mu \nonumber \\
&& \hspace{0.5cm}+\int_{\mathcal{N}_t}\left( D^{q,j,\alpha}(\partial^p_\rho\overline{\phi}_k) \widehat{F}^{q,j,\alpha}_k + D^{q,j,\alpha}(\partial^p_\rho\phi_k) \overline{\widehat{F}}^{q,j,\alpha}_k \right) \mbox{d}\tau \mbox{d}\rho \mbox{d} \mu  \nonumber \\
&& \hspace{0.5cm} + \int_{\mathcal{N}_t}\left(D^{q,j,\alpha}(\partial^p_\rho\overline{\phi}_k) \widehat{G}^{q,j,\alpha}_k + D^{q,j,\alpha}(\partial^p_\rho\phi_k) \overline{\widehat{G}}^{q,j,\alpha}_k\right) \mbox{d}\tau \mbox{d}\rho \mbox{d} \mu  \nonumber \\
&& \hspace{0.5cm}+ \int_{\mathcal{N}_t}\left(D^{q,j,\alpha}(\partial^p_\rho\overline{\phi}_k) \widehat{S}^{q,j,\alpha}_k + D^{q,j,\alpha}(\partial^p_\rho\phi_k) \overline{\widehat{S}}^{q,j,\alpha}_k\right) \mbox{d}\tau \mbox{d} \rho \mbox{d} \mu \nonumber \\
&& \hspace{0.5cm} + \int_{\mathcal{N}_t}\left(D^{q,j,\alpha}(\partial^p_\rho\overline{\phi}_k) \widehat{R}^{q,j,\alpha}_k + D^{q,j,\alpha}(\partial^p_\rho\phi_k) \overline{\widehat{R}}^{q,j,\alpha}_k\right) \mbox{d}\tau \mbox{d} \rho \mbox{d} \mu \nonumber\\
&& \hspace{0.5cm}+ \int_{\mathcal{N}_t}\left(D^{q,j,\alpha}(\partial^p_\rho\overline{\phi}_k) \widehat{U}^{q,j,\alpha}_k + D^{q,j,\alpha}(\partial^p_\rho\phi_k) \overline{\widehat{U}}^{q,j,\alpha}_k\right) \mbox{d}\tau \mbox{d}\rho \mbox{d}\mu \nonumber \\
&& \hspace{0.5cm}+ \int_{\mathcal{N}_t}\left(D^{q,j,\alpha}(\partial^p_\rho\overline{\phi}_k) \widehat{V}^{q,j,\alpha}_k + D^{q,j,\alpha}(\partial^p_\rho\phi_k) \overline{\widehat{V}}^{q,j,\alpha}_k\right)\mbox{d}\tau \mbox{d}\rho \mbox{d} \mu  \nonumber \\
&& \hspace{1cm}\leq \int_{\mathcal{S}_0} \left( |D^{q,j,\alpha}(\partial^p_\rho \phi_k)|^2 + |D^{q,j,\alpha}(\partial^p_\rho \phi_{k+1})|^2\right) \mbox{d}\rho \mbox{d}\mu \nonumber \\ 
&& \hspace{1.5cm} +2\int_{\mathcal{N}_t} \left(p+j-q+k-1-f_k +\left(p-\frac{1}{2}\right)D^{0,1,0}b -\left(q-\frac{1}{2}\right)D^{1,0,0}a\right)|D^{q,j,\alpha}(\partial^p_\rho \phi_k)|^2 \mbox{d} \tau \mbox{d} \rho \mbox{d} \mu, \nonumber
\end{eqnarray*}
for $k=0,1$ and $0\leq j \leq p' \leq m$. As in section
\ref{section:base_step}, the following inequalities will be used:
\begin{eqnarray*}
&& \left|\int_{\mathcal{N}_t} \left( D^{q,j,\alpha}(\partial^p_\rho\overline{\phi}_k) \widehat{F}^{(q,j,\alpha)}_k + D^{q,j,\alpha}(\partial^p_\rho \phi_k ) \overline{\widehat{F}}^{(q,j,\alpha)}_k\right) \mbox{d} \tau \mbox{d} \rho \mbox{d} \mu \right|\leq \int_{\mathcal{N}_t} \left( \varepsilon|D^{q,j,\alpha}(\partial^p_\rho \phi_k)|^2 + \frac{1}{\varepsilon}|\widehat{F}^{(q,j,\alpha)}_k|^2 \right) \mbox{d}\tau \mbox{d} \rho \mbox{d} \mu, \\
&& \left|\int_{\mathcal{N}_t} \left( D^{q,j,\alpha}(\partial^p_\rho \overline{\phi}_k) \widehat{G}^{(q,j,\alpha)}_k + D^{q,j,\alpha}(\partial^p_\rho \phi_k ) \overline{\widehat{G}}^{(q,j,\alpha)}_k\right) \mbox{d} \tau \mbox{d} \rho \mbox{d} \mu  \right| \leq \int_{\mathcal{N}_t} \left( \varepsilon|D^{q,j,\alpha}(\partial^p_\rho \phi_k)|^2 + \frac{1}{\varepsilon}|\widehat{G}^{(q,j,\alpha)}_k|^2 \right) \mbox{d}\tau \mbox{d} \rho \mbox{d} \mu, \\
&& \left|\int_{\mathcal{N}_t} \left( D^{q,j,\alpha}(\partial^p_\rho \overline{\phi}_k) \widehat{U}^{(q,j,\alpha)}_k + D^{q,j,\alpha}(\partial^p_\rho \phi_k ) \overline{\widehat{U}}^{(q,j,\alpha)}_k\right) \mbox{d} \tau \mbox{d} \rho \mbox{d} \mu \right|\leq \int_{\mathcal{N}_t} \left( \zeta|D^{q,j,\alpha}(\partial^p_\rho\phi_k)|^2 + \frac{1}{\zeta}|\widehat{U}^{(q,j,\alpha)}_k|^2 \right) \mbox{d}\tau \mbox{d} \rho \mbox{d} \mu, \\
&& \left|\int_{\mathcal{N}_t} \left( D^{q,j,\alpha}(\partial^p_\rho \overline{\phi}_k) \widehat{V}^{(q,j,\alpha)}_k + D^{q,j,\alpha}(\partial^p_\rho \phi_k ) \overline{\widehat{V}}^{(q,j,\alpha)}_k\right) \mbox{d} \tau \mbox{d} \rho \mbox{d} \mu \right|\leq \int_{\mathcal{N}_t} \left( \zeta|D^{q,j,\alpha}(\partial^p_\rho \phi_k)|^2 + \frac{1}{\zeta}|\widehat{V}^{(q,j,\alpha)}_k|^2 \right) \mbox{d}\tau \mbox{d} \rho \mbox{d} \mu,
\end{eqnarray*}
with $\varepsilon,\zeta>0$. Furthermore, using the same arguments as in
subsection \ref{section:base_step} one has that there is a $\rho_*>0$
such that
\[
|\widehat{S}_k^{(q,j,\alpha)}|^2 \leq \delta, \quad |\widehat{R}_k^{(q,j,\alpha)}|^2 \leq \delta,
\]
for $(\tau,\rho,\varsigma)\in \mathcal{N}_t$, $t \in [0,1]$. Hence,
\begin{eqnarray*}
&& \left|  \int_{\mathcal{N}_t} \left( D^{q,j,\alpha}(\partial^p_\rho\overline{\phi}_k) \widehat{S}_k^{(q,j,\alpha)} + D^{q,j,\alpha}(\partial^p_\rho\phi_k) \overline{\widehat{S}}_k^{(q,j,\alpha)}  \right) \mbox{d}\tau \mbox{d}\rho \mbox{d}\mu  \right| \leq 2\delta \int_{\mathcal{N}_t} \left(1+ |D^{q,j,\alpha} (\partial^p_\rho \phi_k)|^2\right) \mbox{d} \tau \mbox{d} \rho \mbox{d}\mu, \nonumber \\
&& \left|  \int_{\mathcal{N}_t} \left( D^{q,j,\alpha}(\partial^p_\rho\overline{\phi}_{k+1}) \widehat{R}_k^{(q,j,\alpha)} + D^{q,j,\alpha}(\partial^p_\rho\phi_{k+1}) \overline{\widehat{R}}_k^{(q,j,\alpha)}  \right) \mbox{d}\tau \mbox{d}\rho \mbox{d}\mu  \right| \leq 2\delta \int_{\mathcal{N}_t} \left(1+ |D^{q,j,\alpha} (\partial^p_\rho \phi_{k+1})|^2\right) \mbox{d} \tau \mbox{d} \rho \mbox{d}\mu.
\end{eqnarray*}

Using the above estimates one obtains 
\begin{eqnarray*}
&& \int_{\mathcal{S}_t} \bigg( (1+t+a)|D^{q,j,\alpha}(\partial^p_\rho\phi_k)|^2 + (1-t-a) |D^{q,j,\alpha}\partial^p_\rho(\phi_{k+1})|^2 \bigg) \mbox{d}\rho \mbox{d}\mu \nonumber \\
&& \hspace{0.5cm} +\int_{\mathcal{N}_t} (1+c)\left( D^{q,j,\alpha}(\partial^p_\rho \overline{\phi}_k) D^{q,j,\alpha} X_+(\partial^p_\rho\phi_{k+1}) + D^{q,j,\alpha} (\partial^p_\rho\phi_{k+1}) D^{q,j,\alpha}X_+(\partial^p_\rho\overline{\phi}_k) \right) \mbox{d}\tau \mbox{d}\rho \mbox{d}\mu \nonumber \\
&& \hspace{0.5cm} +\int_{\mathcal{N}_t} (1+c)\left(D^{q,j,\alpha}(\partial^p_\rho \phi_k )D^{q,j,\alpha} X_-(\partial^p_\rho\overline{\phi}_{k+1}) + D^{q,j,\alpha}(\partial^p_\rho\overline{\phi}_{k+1}) D^{q,j,\alpha} X_-(\partial^p_\rho \phi_k)\right) \mbox{d}\tau \mbox{d}\rho \mbox{d} \mu \nonumber \\
&& \hspace{0.5cm} +2\int_{\mathcal{N}_t}\left(p+j-q-k-\frac{\varepsilon}{2}-\delta-\frac{\zeta}{2} +g_k+\left(p-\frac{1}{2}\right)D^{0,1,0}b-\left(q-\frac{1}{2}\right)D^{1,0,0}a \right) |D^{q,j,\alpha}(\partial^p_\rho \phi_{k+1})|^2 \mbox{d}\tau \mbox{d} \rho \mbox{d}\mu \nonumber \\
&& \hspace{1cm} \leq \int_{\mathcal{S}_0} \big( |D^{q,j,\alpha}(\partial^p_\rho\phi_k)|^2 + |D^{q,j,\alpha}(\partial^p_\rho\phi_{k+1})|^2\big) \mbox{d} \rho \mbox{d} \mu \nonumber \\
&& \hspace{1.5cm}+ 2 \int_{\mathcal{N}_t} \left(p+j-q+k-1-f_k+\frac{\varepsilon}{2}+\delta+\frac{\zeta}{2}+\left(p-\frac{1}{2}\right)D^{0,1,0}b\right. \nonumber\\
&& \hspace{3cm}\left.-\left(q-\frac{1}{2}\right)D^{1,0,0}a \right)|D^{q,j,\alpha}(\partial^p_\rho \phi_k) |^2 \mbox{d} \tau \mbox{d} \rho \mbox{d} \mu \nonumber \\
&& \hspace{1.5cm} + \int_{\mathcal{N}_t} \left( 2\delta+ \frac{1}{\varepsilon}|\widehat{F}^{(q,j,\alpha)}_k|^2 + \frac{1}{\varepsilon}|\widehat{G}^{(q,j,\alpha)}_k|^2 +\frac{1}{\zeta} |\widehat{U}_k^{(q,j,\alpha)}|^2 +\frac{1}{\zeta} |\widehat{V}_k^{(q,j,\alpha)}|^2 \right) \mbox{d} \tau \mbox{d} \rho \mbox{d} \mu,
\end{eqnarray*}
for a suitable $\rho_*>0$. As in section \ref{section:base_step}, it
is used that $f_k$, $g_k$, $D^{1,0,0}a$ and $D^{0,1,0}b$ are all
$\mathcal{O}(\rho)$. Hence as in the case $j=0$ one can set
\begin{eqnarray*}
&& \left|\left(p-\frac{1}{2}\right)D^{0,1,0}b-\left(q-\frac{1}{2}\right)D^{1,0,0}a    +g_k \right| \leq \eta, \\
&& \left| \left(p-\frac{1}{2}\right)D^{0,1,0}b-\left(q-\frac{1}{2}\right)D^{1,0,0}a  -f_k   \right| \leq \eta,
\end{eqnarray*}
with a suitably small choice of $\rho_*>0$. Putting everything together
one arrives at the inequality
 \begin{eqnarray}
&& \int_{\mathcal{S}_t} \bigg( (1+t+a)|D^{q,j,\alpha}(\partial^p_\rho\phi_k)|^2 + (1-t-a) |D^{q,j,\alpha}\partial^p_\rho(\phi_{k+1})|^2 \bigg) \mbox{d}\rho \mbox{d}\mu \nonumber \\
&& \hspace{0.5cm} +\int_{\mathcal{N}_t} (1+c)\left( D^{q,j,\alpha}(\partial^p_\rho \overline{\phi}_k) D^{q,j,\alpha} X_+(\partial^p_\rho\phi_{k+1}) + D^{q,j,\alpha} (\partial^p_\rho\phi_{k+1}) D^{q,j,\alpha}X_+(\partial^p_\rho\overline{\phi}_k) \right) \mbox{d}\tau \mbox{d}\rho \mbox{d}\mu \nonumber \\
&& \hspace{0.5cm} +\int_{\mathcal{N}_t} (1+c)\left(D^{q,j,\alpha}(\partial^p_\rho \phi_k) D^{q,j,\alpha} X_-(\partial^p_\rho\overline{\phi}_{k+1}) + D^{q,j,\alpha}(\partial^p_\rho\overline{\phi}_{k+1}) D^{q,j,\alpha} X_-(\partial^p_\rho \phi_k)\right) \mbox{d}\tau \mbox{d}\rho \mbox{d} \mu \nonumber \\
&& \hspace{0.5cm} +2\bigg(p+j-q-k-\frac{\varepsilon}{2}-\delta-\frac{\zeta}{2} -\eta\bigg)\int_{\mathcal{N}_t} |D^{q,j,\alpha}(\partial^p_\rho \phi_{k+1})|^2 \mbox{d}\tau \mbox{d} \rho \mbox{d}\mu \nonumber \\
&& \hspace{1.5cm} \leq \int_{\mathcal{S}_0} \big( |D^{q,j,\alpha}(\partial^p_\rho\phi_k)|^2 + |D^{q,j,\alpha}(\partial^p_\rho\phi_{k+1})|^2\big) \mbox{d} \rho \mbox{d} \mu \nonumber \\
&& \hspace{2cm}+ 2\bigg(p+j-q+k-1+\frac{\varepsilon}{2}+\delta+\frac{\zeta}{2}+\eta \bigg) \int_{\mathcal{N}_t} |D^{q,j,\alpha}(\partial^p_\rho \phi_k) |^2 \mbox{d} \tau \mbox{d} \rho \mbox{d} \mu \nonumber \\
&& \hspace{2cm} + \int_{\mathcal{N}_t}  \left( 2\delta+ \frac{1}{\varepsilon}|\widehat{F}^{(q,j,\alpha)}_k|^2 + \frac{1}{\varepsilon}|\widehat{G}^{(q,j,\alpha)}_k|^2 +\frac{1}{\zeta} |\widehat{U}_k^{(q,j,\alpha)}|^2 +\frac{1}{\zeta} |\widehat{V}_k^{(q,j,\alpha)}|^2 \right) \mbox{d} \tau \mbox{d} \rho \mbox{d} \mu, \nonumber \\
&& \label{general_intermediate_integral}
\end{eqnarray} 
for suitably small $\rho_*$ and $k=0,1$. Summing the inequality
(\ref{general_intermediate_integral}) over all admissible values of $q$, $j$
and $\alpha$ for $q+j+|\alpha|\leq m$ and $0\leq j <p'$ one sees that the
terms involving the second and third integral of the left hand side of the
last expression vanish by virtue of lemma \ref{lemma:Liegroups} in appendix \ref{appendix:SU(2)}. Accordingly one has
 \begin{eqnarray}
&&\sum_{q+j+|\alpha|\leq m \atop 0\leq j \leq p'} \int_{\mathcal{S}_t} \bigg( (1+t+a)|D^{q,j,\alpha}(\partial^p_\rho\phi_k)|^2 + (1-t-a) |D^{q,j,\alpha}\partial^p_\rho(\phi_{k+1})|^2 \bigg) \mbox{d}\rho \mbox{d}\mu \nonumber \\
&& \hspace{0.5cm} +2\sum_{q+j+|\alpha|\leq m \atop 0\leq j \leq p'} \bigg(p+j-q-k-\frac{\varepsilon}{2}-\delta-\frac{\zeta}{2} -\eta\bigg)\int_{\mathcal{N}_t} |D^{q,j,\alpha}(\partial^p_\rho \phi_{k+1})|^2 \mbox{d}\tau \mbox{d} \rho \mbox{d}\mu \nonumber \\
&& \hspace{1cm} \leq \sum_{q+j+|\alpha|\leq m \atop 0\leq j \leq p'} \int_{\mathcal{S}_0} \big( |D^{q,j,\alpha}(\partial^p_\rho\phi_k)|^2 + |D^{q,j,\alpha}(\partial^p_\rho\phi_{k+1})|^2\big) \mbox{d} \rho \mbox{d} \mu \nonumber \\
&& \hspace{1.5cm}+ 2\sum_{q+j+|\alpha|\leq m \atop 0\leq j \leq p'} \bigg(p+j-q+k-1+\frac{\varepsilon}{2}+\delta+\frac{\zeta}{2}+\eta \bigg) \int_{\mathcal{N}_t} |D^{q,j,\alpha}(\partial^p_\rho \phi_k) |^2 \mbox{d} \tau \mbox{d} \rho \mbox{d} \mu \nonumber \\
&& \hspace{1.5cm} + \sum_{q+j+|\alpha|\leq m \atop 0\leq j \leq p'} \int_{\mathcal{N}_t}  \left( 2\delta+ \frac{1}{\varepsilon}|\widehat{F}^{(q,j,\alpha)}_k|^2 + \frac{1}{\varepsilon}|\widehat{G}^{(q,j,\alpha)}_k|^2 +\frac{1}{\zeta} |\widehat{U}_k^{(q,j,\alpha)}|^2 +\frac{1}{\zeta} |\widehat{V}_k^{(q,j,\alpha)}|^2 \right) \mbox{d} \tau \mbox{d} \rho \mbox{d} \mu, \nonumber \\
&& \label{general_intermediate_integral_2}
\end{eqnarray} 
for suitable $\rho_*$, $k=0,1$. Now, noting that
\begin{eqnarray*}
&& \sum_{q+j+|\alpha|\leq m \atop 0\leq j \leq p'} \bigg(p+j-q-k-\frac{\varepsilon}{2}-\delta-\frac{\zeta}{2} -\eta\bigg)\int_{\mathcal{N}_t} |D^{q,j,\alpha}(\partial^p_\rho \phi_{k+1})|^2 \mbox{d}\tau \mbox{d} \rho \mbox{d}\mu \nonumber \\
&& \hspace{2cm}  \geq
\bigg(p-m-1-\frac{\varepsilon}{2}-\delta-\frac{\zeta}{2} -\eta\bigg)
\sum_{q+j+|\alpha|\leq m \atop 0\leq j \leq p' } \int_{\mathcal{N}_t} |D^{q,j,\alpha}(\partial^p_\rho \phi_{k+1})|^2 \mbox{d}\tau \mbox{d} \rho \mbox{d}\mu, \\
&& \sum_{q+j+|\alpha|\leq m \atop 0\leq j \leq p'} \bigg(p+j-q+k-1+\frac{\varepsilon}{2}+\delta+\frac{\zeta}{2}+\eta \bigg) \int_{\mathcal{N}_t} |D^{q,j,\alpha}(\partial^p_\rho \phi_k) |^2 \mbox{d} \tau \mbox{d} \rho \mbox{d} \mu \nonumber \\
&& \hspace{2cm} \leq
\bigg(p+m+\frac{\varepsilon}{2}+\delta+\frac{\zeta}{2}+\eta \bigg)
\sum_{q+j+|\alpha|\leq m \atop 0\leq j \leq p'} \int_{\mathcal{N}_t} |D^{q,j,\alpha}(\partial^p_\rho \phi_k) |^2 \mbox{d} \tau \mbox{d} \rho \mbox{d} \mu, 
\end{eqnarray*}
one obtains from (\ref{general_intermediate_integral_2}) the basic inequality
\begin{eqnarray}
&& \sum_{q+j+|\alpha|\leq m \atop 0\leq j \leq p'} \int_{\mathcal{S}_t} \bigg( (1+t+a)|D^{q,j,\alpha}(\partial^p_\rho\phi_k)|^2 + (1-t-a) |D^{q,j,\alpha}\partial^p_\rho(\phi_{k+1})|^2 \bigg) \mbox{d}\rho \mbox{d}\mu \nonumber \\
&& \hspace{0.5cm} +2  \left(p-m-1-\frac{\varepsilon}{2}-\delta
  -\frac{\zeta}{2}-\eta \right) \sum_{q+j+|\alpha|\leq m \atop 0\leq j \leq p'}  \int_{\mathcal{N}_t} |D^{q,j,\alpha}(\partial^p_\rho \phi_{k+1})|^2 \mbox{d}\tau \mbox{d} \rho \mbox{d}\mu \nonumber \\
&& \hspace{1cm} \leq \sum_{q+j+|\alpha|\leq m \atop 0\leq j \leq p'} \int_{\mathcal{S}_0} \big( |D^{q,j,\alpha}(\partial^p_\rho\phi_k)|^2 + |D^{q,j,\alpha}(\partial^p_\rho\phi_{k+1})|^2\big) \mbox{d} \rho \mbox{d} \mu \nonumber \\
&& \hspace{1.5cm}+ 2 \left(p+m+
  \frac{\varepsilon}{2}+\delta+\frac{\zeta}{2}+\eta \right)
\sum_{q+j+|\alpha|\leq m \atop 0\leq j \leq p'}  \int_{\mathcal{N}_t} |D^{q,j,\alpha}(\partial^p_\rho \phi_k) |^2 \mbox{d} \tau \mbox{d} \rho \mbox{d} \mu \nonumber \\
&& \hspace{1.5cm} + \sum_{q+j+|\alpha|\leq m \atop 0\leq j \leq p'} \int_{\mathcal{N}_t}  \left( 2\delta+ \frac{1}{\varepsilon}|\widehat{F}^{(q,j,\alpha)}_k|^2 + \frac{1}{\varepsilon}|\widehat{G}^{(q,j,\alpha)}_k|^2 +\frac{1}{\zeta} |\widehat{U}_k^{(q,j,\alpha)}|^2 +\frac{1}{\zeta} |\widehat{V}_k^{(q,j,\alpha)}|^2 \right)  \mbox{d} \tau \mbox{d} \rho \mbox{d} \mu, \nonumber
\\ \label{general_basic_inequality}
\end{eqnarray} 
for a suitably small $\rho_*>0$ and $k=0,1$. As in the discussion of the
preceding sections, the integers $p$ and $m$ are to be chosen such that the
second term in inequality (\ref{general_basic_inequality}) is
positive. That is, one requires
\[
p > m+1+\frac{\varepsilon}{2}+\delta +\frac{\zeta}{2}+\eta.
\] 
The constants $\varepsilon$, $\delta$, $\zeta$, $\eta$ can be chosen such that
\[
\frac{\varepsilon}{2}+\delta +\frac{\zeta}{2}+\eta <1,
\] 
by means of a suitably small choice of $\rho_*$. Note that the
``shrinking'' of $\rho_*$ is only performed a finite number of times,
hence the neighbourhood around spatial infinity does not collapse to a
point. Using the same arguments leading to inequality
(\ref{schw_ineq_2}) one obtains from (\ref{general_basic_inequality})
that
\begin{eqnarray}
&& \sum_{q+j+|\alpha|\leq m \atop 0\leq j \leq p'} \int_{\mathcal{S}_t} |D^{q,j,\alpha}(\partial^p_\rho\phi_k)|^2  \mbox{d}\rho \mbox{d}\mu  \nonumber \\
&& \hspace{1cm}\leq \sum_{q+j+|\alpha|\leq m \atop 0\leq j \leq p'} \int_{\mathcal{S}_0} \big( |D^{q,j,\alpha}(\partial^p_\rho\phi_k)|^2 + |D^{q,j,\alpha}(\partial^p_\rho\phi_{k+1})|^2\big) \mbox{d} \rho \mbox{d} \mu \nonumber \\
&& \hspace{1.5cm}+ 2 (p+m+ 1) \sum_{q+j+|\alpha|\leq m \atop 0\leq j \leq p'}  \int_{\mathcal{N}_t} |D^{q,j,\alpha}(\partial^p_\rho \phi_k) |^2 \mbox{d} \tau \mbox{d} \rho \mbox{d} \mu \nonumber \\
&& \hspace{1.5cm} + \sum_{q+j+|\alpha|\leq m \atop 0\leq j \leq p'} \int_{\mathcal{N}_t}  \left( 2\delta+ \frac{1}{\varepsilon}|\widehat{F}^{(q,j,\alpha)}_k|^2 + \frac{1}{\varepsilon}|\widehat{G}^{(q,j,\alpha)}_k|^2 +\frac{1}{\zeta} |\widehat{U}_k^{(q,j,\alpha)}|^2 +\frac{1}{\zeta} |\widehat{V}_k^{(q,j,\alpha)}|^2 \right)    \mbox{d} \tau \mbox{d} \rho \mbox{d} \mu, \nonumber \\
&& \label{general_reduced_inequality}
\end{eqnarray} 
valid for $t\in[0,1]$, $p\geq m+2$, $k=0,1$, and suitably small $0<\rho_* <a_*$.
The companion inequality for $\phi_{2}$ is
\begin{eqnarray}
&& 2 (p-m-2) \sum_{q+j+|\alpha|\leq m \atop 0\leq j \leq p'}  \int_{\mathcal{N}_t} |D^{q,j,\alpha}(\partial^p_\rho \phi_{2})|^2 \mbox{d}\tau \mbox{d} \rho \mbox{d}\mu \nonumber \\
&& \hspace{1cm} \leq \sum_{q+j+|\alpha|\leq m \atop 0\leq j \leq p'} \int_{\mathcal{S}_0} \big( |D^{q,j,\alpha}(\partial^p_\rho\phi_1)|^2 + |D^{q,j,\alpha}(\partial^p_\rho\phi_{2})|^2\big) \mbox{d} \rho \mbox{d} \mu \nonumber \\
&& \hspace{1.5cm}+ 2(p+m+1) \sum_{q+j+|\alpha|\leq m \atop 0\leq j \leq p'}  \int_{\mathcal{N}_t} |D^{q,j,\alpha}(\partial^p_\rho \phi_1) |^2 \mbox{d} \tau \mbox{d} \rho \mbox{d} \mu \nonumber \\
&& \hspace{1.5cm} + \sum_{q+j+|\alpha|\leq m \atop 0\leq j \leq p'} \int_{\mathcal{N}_t}  \left( 2\delta+ \frac{1}{\varepsilon}|\widehat{F}^{(q,j,\alpha)}_1|^2 + \frac{1}{\varepsilon}|\widehat{G}^{(q,j,\alpha)}_1|^2 +\frac{1}{\zeta} |\widehat{U}_1^{(q,j,\alpha)}|^2 +\frac{1}{\zeta} |\widehat{V}_1^{(q,j,\alpha)}|^2 \right)    \mbox{d} \tau \mbox{d} \rho \mbox{d} \mu, \nonumber \\
&& \label{companion_general_reduced_inequality}
\end{eqnarray}  
for $t\in[0,1]$, $p\geq m+2$ and suitably small $\rho_*$. Gronwall's
argument applied to (\ref{general_reduced_inequality}) yields
\begin{eqnarray}
&&  \hspace{-2.2cm}\sum_{q+j+|\alpha|\leq m \atop 0\leq j \leq p'} \int_{\mathcal{N}_t} |D^{q,j,\alpha}(\partial^p_\rho\phi_k)|^2  \mbox{d}\tau \mbox{d}\rho \mbox{d}\mu \nonumber \\
&& \hspace{-1.8cm} \leq\frac{1}{2(p+m+1)}\left( e^{2(p+m+1)t}-1 \right)
\sum_{q+j+|\alpha|\leq m \atop 0\leq j \leq p'} \int_{\mathcal{S}_0} \big( |D^{q,j,\alpha}(\partial^p_\rho\phi_k|^2) + |D^{q,j,\alpha}(\partial^p_\rho\phi_{k+1})|^2\big) \mbox{d} \rho \mbox{d} \mu \nonumber \\
&&  \hspace{-1.3cm} +e^{2(p+m+1)t} \int^t_0 e^{-2(p+m+1)s} \left(
  \sum_{q+j+|\alpha|\leq m \atop 0\leq j \leq p'} \int_{\mathcal{N}_t} \left( 2\delta+ \frac{1}{\varepsilon}|\widehat{F}^{(q,j,\alpha)}_k|^2 + \frac{1}{\varepsilon}|\widehat{G}^{(q,j,\alpha)}_k|^2 +\frac{1}{\zeta} |\widehat{U}_k^{(q,j,\alpha)}|^2 +\frac{1}{\zeta} |\widehat{V}_k^{(q,j,\alpha)}|^2 \right) \mbox{d} \tau \mbox{d} \rho \mbox{d} \mu \right)   \mbox{d} s, \nonumber \\
&& \label{general_gronwall}
\end{eqnarray}
for $k=0,1$, $t\in[0,1]$ and suitably small $0<\rho_*<a_*$. To
conclude the argument it is noted that because of the induction
hypothesis, there are constants $\hat{C}_1(\rho_*)>0$ and
$\hat{C}_2(\rho_*)>0$, such that
\begin{eqnarray}
&&\sum_{q+j+|\alpha|\leq m \atop 0\leq j \leq p'} \int_{\mathcal{N}_t}
\left( |\hat{U}_k^{(q,j,\alpha)}|^2 + |\hat{V}_k^{(q,j,\alpha)}|^2 \right)
 \mbox{d}\rho \mbox{d}\mu \nonumber \\ 
&& \hspace{2cm}\leq
\hat{C}_1(\rho_*) \sum_{k=0}^2
\int_{\mathcal{S}_0} \left(\sum_{q+j+|\alpha|\leq m \atop 0\leq j \leq p'}|D^{(q,j,\alpha)}(\partial^p_\rho \phi_k)|^2 \right)\mbox{d}
\tau \mbox{d} \rho \mbox{d}\mu +  \hat{C}_2(\rho_*). \label{explicit_ind_hyp} 
\end{eqnarray}
Furthermore, if the solutions,  to the transport equations on $\mathcal{I}$, $\phi_k^{(l)}$ for $k=0,1,2$ and $0\leq l \leq
p$, are suitably smooth ---and this can be controlled by a suitable choice of
the initial data--- then there is a constant $\hat{D}(\rho_*)>0$ such that  
\begin{equation}
\sum_{q+j+|\alpha|\leq m \atop 0\leq j \leq p'} \int_{\mathcal{N}_t}
\left( |\hat{F}_k^{(q,j,\alpha)}|^2 + |\hat{G}_k^{(q,j,\alpha)}|^2 \right)
\mbox{d}\tau \mbox{d}\rho \mbox{d}\mu \leq \hat{D}(\rho_*). \label{cylinder_estimate}
\end{equation}
Thus, combining the inequalities (\ref{general_reduced_inequality}) and
(\ref{companion_general_reduced_inequality}) ---following the model of section
\ref{section:base_step}--- and using (\ref{explicit_ind_hyp}) and
(\ref{cylinder_estimate}) one finds that there exist constants
$\tilde{C}_1(\rho_*)>0$ and $\tilde{C}_2(\rho_*)>0$ such that
\begin{eqnarray*}
&&  \int_{\mathcal{N}_t} \sum_{q+j+|\alpha|\leq m \atop 0\leq j \leq p'}
|D^{q,j,\alpha}(\partial^p_\rho \phi_k)|^2 \mbox{d}\tau \mbox{d}\rho \mbox{d}
\mu \nonumber \\
&&\hspace{2cm} \leq \tilde{C}_1(\rho_*) 
\sum^2_{k=0} \int_{\mathcal{S}_0} \left( \sum_{q+j+|\alpha|\leq m \atop 0 \leq
    j \leq p'}
  |D^{q,j,\alpha}(\partial^p_\rho \phi_k)|^2 \right) 
\mbox{d}\rho \mbox{d} \mu + \tilde{C}_2(\rho_*),  
\end{eqnarray*}
for suitably small $\rho_*>0$ and $p>m+2$. This concludes the
induction argument.

\section{Conclusions} \label{section:conclusions}
Sections \ref{section:base_step} and \ref{section:general_step} have
been concerned with the construction of $L^2$-type estimates for the
quantities $\partial^p_\rho \phi_k$, $k=0,1,2$ in the case of a
Maxwell field propagating as a test field on a Schwarzschild
background. The basic assumptions and main results of this construction
are summarised in the following theorem.

\begin{theorem} \label{theorem:main}
Let $m>0$ an integer and $\phi_k$, $k=0,1,2$ solutions to the
Maxwell equations on a Schwarzschild background, equations (\ref{A0})
to (\ref{B2}). Furthermore, assume $\phi_k$ $k=0,1,2$ are of the form 
\[
\phi_k = \sum_{l=|1-k|}^{p-1} \frac{1}{l!} \phi^{(l)}_k \rho^l +
J^p(\partial^p_\rho \phi_k),
\]
with $\phi^{(l)}_k$, $0\leq l \leq p-1$ solutions of the transport
equations implied by the Maxwell equations on $\mathcal{I}$. Assume
that the initial data for the equations (\ref{A0}) to (\ref{B2}) are 
such that the functions $\phi^{(l)}_k$ are suitably smooth ---so that
that the functions $\hat{F}^{(q,j,\alpha)}_k$ and
$\hat{G}^{(q,j,\alpha)}_k$ are bounded in a neighbourhood of
$\mathcal{I}$. Then, given $p>m+2$ there exist  $0 <\rho_* <
a_*$ and constants $C_1(\rho_*)$, $C_2(\rho_*)$ such that
\begin{eqnarray*}
&&  \int_{\mathcal{N}_t} \sum_{q+j+|\alpha|\leq m}
|D^{q,j,\alpha}(\partial^p_\rho \phi_k)|^2 \mbox{d}\tau \mbox{d}\rho \mbox{d}
\mu \nonumber \\
&&\hspace{2cm} \leq C_1(\rho_*) 
\sum^2_{k=0} \int_{\mathcal{S}_0} \left( \sum_{q+j+|\alpha|\leq m}
  |D^{q,j,\alpha}(\partial^p_\rho \phi_k)|^2 \right) 
\mbox{d}\rho \mbox{d} \mu + C_2(\rho_*),  
\end{eqnarray*}
for $t\in[0,1]$.
\end{theorem}

\textbf{Remark.} The precise nature of the regularity of the solutions
to the transport equations on $\mathcal{I}$ must be analysed on a case
by case basis, and can be obtained by means of a careful reading of the
arguments presented in this article. Their precise formulation goes beyond
the scope of the present work. A translation of the conditions in
terms of initial data can be obtained by means of the techniques of
\cite{Val07b}.

\bigskip
To conclude, assuming that one has a solution of the Maxwell equations
on a Schwarzschild background of the form (\ref{Ansatz}), then the
estimates in theroem \ref{theorem:main} allow to control the
regularity of the remainder in the expansion in a similar way as it
was done in section \ref{section:Minkowski} for a flat background, at
least for a small (but finite) neighbourhood of the cylinder at
spatial infinity, $\mathcal{I}$, which includes the critical set,
$\mathcal{I}^+$, and future null infinity, $\scri^+$. The open
question is now how to ensure the existence of solutions to the
Maxwell equations of the desired form.

\section{Acknowledgments}
This research is funded by an EPSRC Advanced Research Fellowship. I
thank CM Losert-Valiente Kroon for a careful reading of the manuscript.

\appendix

\section{Some spinors} \label{appendix:spinors}
Let $\{o_A, \iota_A\}$ denote a normalised spinor dyad,
$\epsilon_{AB}o^A\iota^B=1$, where $\epsilon_{AB}$ is the standard alternating
spinor. The following spinors have been used:
\begin{subequations}
\begin{eqnarray}
&& \tau_{AA'}=o_A o_{A'} +\iota_A \iota_{A'}, \\
&& \epsilon^0_{AB}=o_A o_B, \quad \epsilon^1_{AB}=o_{(A}\iota_{B)}, \quad \epsilon^2_{AB}=\iota_A \iota_B, \\
&& x_{AB}=\frac{1}{\sqrt{2}}(o_A\iota_B+\iota_A o_B), \quad y_{AB}=-\frac{1}{\sqrt{2}}\iota_A\iota_B, \quad z_{AB}=\frac{1}{\sqrt{2}}o_A o_B, \\
&& \epsilon^0_{ABCD}=o_A o_B o_C o_D, \quad \epsilon^1_{ABCD}= o_{(A} o_B o_C \iota_{D)}, \quad \cdots, \quad \epsilon^4_{ABCD}=\iota_A \iota_B \iota_C \iota_D, \\
&& h_{ABCD}=-\epsilon_{A(C}\epsilon_{D)B}.
\end{eqnarray}
\end{subequations}

\section{Some commutators on $SU(2)$} \label{appendix:SU(2)}
Consider the basis 
\[
u_1=\frac{1}{2} \begin{pmatrix} 0 & \mbox{i} \\ \mbox{i} & 0 \end{pmatrix}, \quad
u_2=\frac{1}{2} \begin{pmatrix} 0 & -1 \\ 1 & 0 \end{pmatrix}, \quad
u_3=\frac{1}{2} \begin{pmatrix} \mbox{i} & 0 \\ 0 & -\mbox{i} \end{pmatrix},
\]
of the Lie algebra $su(2)$, with $u_3$ the generator of the subgroup
$U(1)$. Let $Z_i$, $i=1,2,3$ denote the (real) left invariant vector
fields generated by $u_i$ on the Lie group $SU(2)$. They satisfy the
following commutator relations
\begin{equation}
[Z_1,Z_2]=Z_3, \quad [Z_1,Z_3]=-Z_2, \quad [Z_2,Z_3]=Z_1.
\end{equation}
We also make use of the following combinations of the $Z_i$ 
\begin{equation}
X_+=-Z_2-\mbox{i}Z_1, \quad X_-=-Z_2+\mbox{i}Z_1, \quad X=-2\mbox{i}Z_3.
\end{equation}
Their commutators are
\begin{equation}
[X,X_+]=2X_+, \quad [X,X_-]=-2X_-, \quad [X_+,X_-]=-X.
\end{equation}

As in the main text let $Z^\alpha=Z_1^{\alpha_1} Z_2^{\alpha_2}
Z_3^{\alpha_3}$. The operators provide a basis for the universal
enveloping algebra of $su(2)$. The following result of \cite{Fri03b}
has been used several times in the main text.

\begin{lemma} \label{lemma:Liegroups}
For any smooth complex-valued functions $f$, $g$ on $SU(2)$ the
operators $Z^\alpha$ satisfy
\begin{equation}
\sum_{|\alpha|\leq m} \int_{SU(2)} (Z^\alpha X_\pm f Z^\alpha g + Z^\alpha f Z^\alpha X_\pm g) \mbox{d}\mu =0,
\end{equation}
where $\mbox{d}\mu$ denotes the normalised Haar measure on $SU(2)$.
\end{lemma}


\begin{thebibliography}{10}

\bibitem{Fri95}
H.~Friedrich,
\newblock {\em {Einstein} equations and conformal structure: existence of
  anti-de {Sitter}-type space-times},
\newblock J. Geom. Phys. {\bf 17}, 125 (1995).

\bibitem{Fri98a}
H.~Friedrich,
\newblock {\em Gravitational fields near space-like and null infinity},
\newblock J. Geom. Phys. {\bf 24}, 83 (1998).

\bibitem{Fri03a}
H.~Friedrich,
\newblock {\em Conformal Einstein evolution},
\newblock in {\em The conformal structure of spacetime: Geometry, Analysis,
  Numerics}, edited by J.~Frauendiener \& H.~Friedrich, Lecture Notes in
  Physics, page~1, Springer, 2002.

\bibitem{Fri03b}
H.~Friedrich,
\newblock {\em Spin-2 fields on Minkowski space near space-like and null
  infinity},
\newblock Class. Quantum Grav. {\bf 20}, 101 (2003).

\bibitem{Fri04}
H.~Friedrich,
\newblock {\em Smoothness at null infinity and the structure of initial data},
\newblock in {\em 50 years of the Cauchy problem in general relativity}, edited
  by P.~T. Chru\'{s}ciel \& H.~Friedrich, Birkhausser, 2004.

\bibitem{Joh91}
F.~John,
\newblock {\em Partial differential equations},
\newblock Springer, 1991.

\bibitem{Som80}
P.~Sommers,
\newblock {\em Space spinors},
\newblock J. Math. Phys. {\bf 21}, 2567 (1980).

\bibitem{Ste91}
J.~Stewart,
\newblock {\em Advanced general relativity},
\newblock Cambridge University Press, 1991.

\bibitem{Tay96}
M.~E. Taylor,
\newblock {\em Partial differential equations {I}},
\newblock Springer, 1996.

\bibitem{Val03a}
J.~A. Valiente~Kroon,
\newblock {\em Polyhomogeneous expansions close to null and spatial infinity},
\newblock in {\em The Conformal Structure of Spacetimes: Geometry, Numerics,
  Analysis}, edited by J.~Frauendiner \& H.~Friedrich, Lecture Notes in
  Physics, page 135, Springer, 2002.

\bibitem{Val04d}
J.~A. Valiente~Kroon,
\newblock {\em Does asymptotic simplicity allow for radiation near spatial
  infinity?},
\newblock Comm. Math. Phys. {\bf 251} (2004).

\bibitem{Val04a}
J.~A. Valiente~Kroon,
\newblock {\em A new class of obstructions to the smoothness of null infinity},
\newblock Comm. Math. Phys. {\bf 244}, 133 (2004).

\bibitem{Val04e}
J.~A. Valiente~Kroon,
\newblock {\em Time asymmetric spacetimes near null and spatial infinity. I.
  Expansions of developments of conformally flat data},
\newblock Class. Quantum Grav. {\bf 23}, 5457 (2004).

\bibitem{Val05a}
J.~A. Valiente~Kroon,
\newblock {\em Time asymmetric spacetimes near null and spatial infinity. II.
  Expansions of developments of initial data sets with non-smooth conformal
  metrics},
\newblock Class. Quantum Grav. {\bf 22}, 1683 (2005).

\bibitem{Val07b}
J.~A. Valiente~Kroon,
\newblock {\em The Maxwell field on the Schwarzschild spacetime: behaviour near
  spatial infinity},
\newblock Proc. Roy. Soc. Lond. A {\bf 463}, 2609 (2007).

\end{thebibliography}

\end{document}